\documentclass[12pt,a4paper]{article}

\usepackage{jheppub}

\usepackage[latin1]{inputenc}
\usepackage{amsmath}

\DeclareMathOperator{\Erf}{Erf}

\DeclareMathOperator\Erfi{Erfi}

\usepackage{amsfonts}
\usepackage{physics}
\usepackage{amssymb}
\usepackage{faktor}
\usepackage{graphicx}
\usepackage{tikz}
\usepackage{amsthm}
\numberwithin{equation}{section}
\numberwithin{figure}{section}
\usepackage[english]{babel}
\usepackage{mathtools}
\usepackage{nicefrac}
\usepackage{lscape}
\usepackage{textgreek}
\usepackage{parskip}
\usepackage{enumerate}

\usepackage{tcolorbox}

\usepackage{todonotes}

\usepackage{dsfont}

\usepackage{url}

\newtheorem{theorem}{Theorem}[section]

\theoremstyle{definition}

\usepackage{subcaption}
 \usepackage{cleveref}

\hypersetup{
	pdftitle={Unorientable WP Volumes},
	pdfauthor={Torsten Weber}}
\usepackage{upgreek}
\usepackage{array}
\usepackage{bm}

\author[a]{Torsten Weber,}
\author[b]{Jarod Tall,}
\author[a]{Fabian Haneder,}
\author[a]{Juan Diego Urbina,}
\author[a]{and Klaus Richter}
\affiliation[a]{Institut f\"ur Theoretische Physik, 
Universit\"at Regensburg, \\
Universit\"atsstr. 31, D-93040 Regensburg, Germany}
\affiliation[b]{Department of Physics and Astronomy, Washington State University,\\
Pullman, WA 99164-2814 USA}

\emailAdd{torsten.weber@ur.de}
\emailAdd{jarod.tall@wsu.edu}
\emailAdd{fabian.haneder@ur.de}
\emailAdd{juan-diego.urbina@ur.de}
\emailAdd{klaus.richter@ur.de}
\title{Unorientable topological gravity and orthogonal random matrix universality}

\abstract{
The duality of Jackiw-Teitelboim (JT) gravity and a double scaled matrix integral has led to studies of the canonical spectral form factor (SFF) in the so called $\tau-$scaled limit of large times, $t \to \infty$, and fixed temperature in order to demonstrate agreement with universal random matrix theory (RMT). Though this has been established for the unitary case, extensions to other symmetry classes requires the inclusion of unorientable manifolds in the sum over geometries, necessary to address time reversal invariance, and regularization of the corresponding prime geometrical objects, the Weil-Petersson (WP) volumes. We report here how universal signatures of quantum chaos, witnessed by the fidelity to the Gaussian orthogonal ensemble, emerge for the low-energy limit of unorientable JT gravity, i.e. the unorientable Airy model/topological gravity. To this end, we implement the loop equations for the corresponding dual (double-scaled) matrix model and find the generic form of the unorientable Airy WP volumes, supported by calculations using unorientable Kontsevich graphs. In an apparent violation of the gravity/chaos duality, the $\tau-$scaled SFF on the gravity side acquires both logarithmic and power law contributions in $t$, not manifestly present on the RMT side. We show the expressions can be made to agree by means of bootstrapping-like relations hidden in the asymptotic expansions of generalized hypergeometric functions. Thus, we are able to establish strong evidence of the quantum chaotic nature of unorientable topological gravity. 

}

\makeindex

\begin{document}
\maketitle

\section{Introduction}
\label{sec:intro}
The recent work demonstrating the duality between the path integral of Jackiw-Teitelboim (JT) gravity \cite{Jackiw1985} and a certain double scaled matrix integral \cite{Saad2019, Stanford2019} has allowed for direct calculations of $n$-point correlation functions in terms of a perturbative genus expansion, making JT gravity a useful toy model for studying quantum gravity. The link to a matrix integral strongly suggests the notion that JT gravity behaves as a quantum chaotic system. Indeed, the very definition of quantum chaos, as conjectured by Bohigas, Giannoni and Schmit \cite{Bohigas1984}, is that the spectral statistics of a quantum Hamiltonian should be described by random matrix theory (RMT) in the proper universal regime. The agreement between the spectral statistics of JT gravity dual to a unitary matrix model and universal RMT has recently been established analytically \cite{Saad2022}. Furthermore, in the low temperature limit, JT gravity is described by the same Schwarzian action as the SYK model \cite{Kitaev2016_1,Kitaev2016_2}, and numerical evidence has established agreement between the spectral statistics of the SYK model and RMT universality \cite{Cotler2017}. The universal chaos bound \cite{Maldacena2015}, calculated from out-of-time-order-correlators, has also been shown to be saturated, both analytically and numerically, for the SYK model \cite{Maldacena2016b, Jensen2016, Kobrin2020}. However, there has been a lack of exploration of the quantum chaotic properties of unorientable JT gravity (corresponding to the time-reversal invariant case) due to the theory being divergent \cite{Stanford2019}.

To be precise, one can consider the canonical spectral form factor (SFF), \\
$\ev{ Z(\beta +it)Z(\beta-it)}_c$, as a useful diagnostic for chaos in quantum systems \cite{Saad2018}. The quantity can be computed from an analytic continuation of the connected two-point function,  $\ev{ Z(\beta_1)Z(\beta_2)}_c$. The path integral gives an interpretation of $\ev{ Z(\beta_1)Z(\beta_2)}_c$ as a sum over all connected spacetime topologies with two asymptotic boundaries, and the solution can be written as a genus expansion in the parameter $e^{S_0}$. The parameter $S_0$ is a coupling in the gravitational action and can be interpreted as the leading order contribution to the entropy \cite{Johnson:2020exp}. The important point is that the path integral calculation reduces to computing the volume of the moduli space of the topologies under consideration. For example, when the moduli space consists of bordered orientable Riemann surfaces, as shown in \cite{Saad2019}, the volume of the moduli space is given by the Weil-Petersson (WP) volume, which can be computed effectively from Mirzakhani's recursion relation \cite{Mirzakhani2006, Mirzakhani2007}. In general, the manifolds included in the path integral can be orientable or unorientable, contain additional spin or pin structures, or they can be supersymmetric. The precise specification determines the symmetry class of the dual matrix model. The complete classification of such variations of JT gravity and their dual matrix model theories was done in \cite{Stanford2019}. 

The focus of this paper is the case
when the manifolds are allowed to be unorientable, but contain no additional structures and have no supersymmetry, corresponding to a matrix model in the orthogonal symmetry class. This can be considered as a more important case than the unitary symmetry class because it implies time reversal invariance of the corresponding boundary theory. The necessary generalization of the WP volumes to unorientable surfaces contains logarithmic divergences and the volume is therefore infinite \cite{Norbury2008, Gendulphe2017}. However, it was recently shown by utilizing a regularization scheme that the unorientable WP volumes satisfy a Mirzakhani type recursion relation \cite{Stanford2023}. This unorientable Mirzakhani recursion relation is then related via Laplace transform to the standard loop equations \cite{Eynard2018, Stanford2019} of a matrix model with orthogonal symmetry and the JT gravity leading order energy density $\rho^{\text{JT}}_0(E) \propto \sinh(2\pi \sqrt{E})$. In the low energy limit JT gravity reduces to the Airy model, also known as topological gravity, with energy density $\rho^{\text{Airy}}_0(E) \propto \sqrt{E}$. For this model, the volume of the moduli space is finite, allowing one to define the unorientable Airy WP volumes without any issue.

The main task of this work is to compute the canonical SFF of unorientable topological gravity and compare the result to universal random matrix theory. On the gravitational side it is necessary to first compute the unorientable Airy WP volumes for the case of two boundaries. To do so we will use the loop equations with the Airy spectral curve. The universal random matrix theory result of the canonical SFF can be derived from the Laplace transform of the form factor, which is well established in the quantum chaos literature \cite{Mehta2004, Haake2010}. The form factor in the universal regime depends only on the symmetry class and mean level spacing of the matrix model. In a traditional, i.e. not double scaled, matrix model the universal limit is $N \to \infty$, where $N$ is the dimensionality of the matrices. In practice, for a double scaled matrix model $N$ will be replaced by $e^{S_0}$, and the relevant limit will be $e^{S_0} \to \infty$. This quantity can be interpreted as the typical value of the eigenvalue density of the matrix model in this double scaled limit. 

For the unorientable Airy model to be quantum chaotic, the behavior of the canonical SFF should agree with the universal RMT result for the orthogonal symmetry class in the late time limit, see \cite{Altland2020} for a discussion on time scales in quantum chaos. This time, $t$, is taken to be on the order of $e^{S_0}$, i.e. $t \sim e^{S_0}$. We will work in the so called ``$\tau-$scaling'' limit where both $e^{S_0} \to \infty$ and $ t \to \infty$ but the quantity $\tau \coloneqq t e^{-S_0}$ is held fixed. This limit has been studied somewhat extensively \cite{Saad2022,Okuyama2020a, Okuyama2021, Okuyama2023} for the orientable Airy model corresponding to the unitary symmetry class. In this case, it is straightforward to show the canonical SFF of the Airy model agrees with universal RMT after $\tau-$scaling. However, it will be shown the canonical SFF of the unorientable Airy model has terms that are higher order in $e^{S_0}$, so that powers of $t$ and terms logarithmic in $t$ survive the $\tau-$scaling limit. The universal RMT result, by its very definition, is independent of $t$ after $\tau-$scaling, and thus the comparison between the results becomes fairly non-trivial. In order to reconcile the two expressions we use asymptotic expansions of generalized hypergeometric functions to derive an identity, valid in the limit $t \to \infty$, to demonstrate equivalence of the expressions for low orders in $\tau$. We then argue how similar identities could be derived to demonstrate equivalence for higher order terms.

We now present necessary background material and give a summary of the main results of the paper.
\subsection{Canonical SFF from unorientable topological gravity}
The connected two-point function of unorientable topological gravity has the following genus expansion \cite{Stanford2019}:
\begin{align}
    \ev{Z(\beta_1)Z(\beta_2)}_c = \sum_{g=0,\frac{1}{2},1 \dots}\frac{Z_{g,2}(\beta_1,\beta_2)}{\left(e^{S_0}\right)^{2g}},
\end{align}
where
\begin{align}
    Z_{g,2}(\beta_1,\beta_2)=\int_0^\infty \int_0^\infty b_1\dd{b_1} b_2\dd{b_2}Z^t(b_1,\beta_1) Z^t(b_2,\beta_2)V^{\text{Airy}}_{g,2}(b_1,b_2),
\end{align}
with the trumpet partition function defined as
\begin{align}
    Z^t(b,\beta)\coloneqq\frac{1}{\sqrt{4\pi \beta}}e^{-\frac{b^2}{4\beta}},
\end{align}
and $V^{\text{Airy}}_{g,2}$ are the unorientable Airy WP volumes. The SFF is found by taking $\beta_1 = \beta +i t$ and $\beta_2 = \beta -i t$, i.e.
\begin{align}
    \kappa_{\beta}(t) \coloneqq \ev{Z(\beta+it)Z(\beta-it)}_c = \sum_{g=0,\frac{1}{2},1 \dots}\frac{\kappa^g_{\beta}(\beta_1,\beta_2)}{\left(e^{S_0}\right)^{2g}}.\label{eq:sff_def}
\end{align}
Therefore, once the $V^{\text{Airy}}_{g,2}$ are known, computing the SFF is straightforward. The easiest way to compute the volumes is to take advantage of the duality to a double scaled matrix integral of orthogonal symmetry class. The duality is defined by the following correspondence \cite{Eynard2007, Stanford2023}:
\begin{align}
    V^{\text{Airy}}_{g,n}(b_1,\dots,b_n)&=\mathcal{L}^{-1}\qty[\prod_{i=1}^n\qty( \frac{-2z_i}{b_i})R^{\text{Airy}}_{g,n}(-z_1^2,\dots,-z_n^2),\qty(b_1,\dots,b_n)] \\
    &=(-1)^n\int_{\delta +i\mathbb{R}}R^{\text{Airy}}_{g,n}\left(-z_1^2,\dots,-z_n^2\right)\prod_{j=1}^n \frac{dz_j}{2\pi i}\frac{2z_j}{b_j}e^{b_j z_j} 
    \label{eq:volume}
\end{align}
The $R_{g,n}$ are the coefficients of the genus expansion of the connected $n-$point correlation function of the resolvents taken from a double scaled matrix integral of orthogonal symmetry class. Here we will mainly be interested in the case $n=2$. The Airy limit of the WP volumes is given by the limit $b_1, \dots, b_n \to \infty$. This limit can be computed using~\eqref{eq:volume} by taking the leading order energy density of the matrix model to be:
\begin{align}
    \rho^{\text{Airy}}_0(E) = \frac{\sqrt{E}}{2\pi}.
    \label{eq:ed}
\end{align}
With this information the $R^{\text{Airy}}_{g,n}$ can be recursively computed using the loop equations \cite{Eynard2018,Stanford2019}. In this work we compute $R^{\text{Airy}}_{g,2}$, and correspondingly $V^{\text{Airy}}_{g,2}$, up to $g=7/2$. The unorientable Airy WP volumes we find are of the form 
\begin{align}
    V^{\text{Airy}}_{g,2}(b_1,b_2)=V^>_g(b_1,b_2)\theta\qty(b_1-b_2)+V^>_g(b_2,b_1)\theta\qty(b_2-b_1),
\end{align}
with 
\begin{align}
	V^{>}_{g}(b_1,b_2)=\sum_{\alpha_1,\alpha_2 \in \mathbb{N}_0}^{\alpha_1+\alpha_2= 6g-2}C_{\alpha_1,\alpha_2}b_1^{\alpha_1}b_2^{\alpha_2},
\end{align}
and in general not symmetric coefficients $C_{\alpha_1,\alpha_2}$.
The contributions to \eqref{eq:sff_def} up to $g=7/2$ derived from these volumes are reported in section \ref{sec:AiryWP}. Here we report the first few terms after taking the $\tau-$scaling limit:
\begin{align}
   e^{-S_0}\kappa^{\text{WP}}_{\beta}(\tau) &= \frac{\tau}{2\pi \beta} -\frac{\tau^2}{\sqrt{2\pi \beta }}+ \frac{\tau^3}{\pi} \qty[\frac{-10}{3}+\log\qty(\frac{2t}{\beta})-\frac{\sqrt{2\pi}}{3}\left(t\tau^2\right)^{1/2} -\frac{2\left(t\tau^2\right)^2}{45}]\nonumber\\[5pt] 
 &+\frac{8\sqrt{2\pi \beta}}{3} \tau^4 +\order{\tau^5},
 \label{eq:19}
\end{align}
where the superscript WP was added to indicate this was computed from the unorientable Airy WP volumes as opposed to universal RMT. Note that powers of $t$ and logarithmic terms in $t$ remain after $\tau-$scaling. It will become clear why the coefficient of $\tau^3$ is presented in this way after comparing to the SFF computed from universal RMT.
\subsection{Canonical SFF from universal RMT with orthogonal symmetry}
The canonical SFF of universal RMT, after $\tau-$scaling, is given by:
\begin{align}
  e^{-S_0}\kappa_{\beta}(\tau) &=\int_{0}^{\infty}\dd E e^{-2\beta E}\rho_0(E)- 
\int_{0}^{\infty}\dd E e^{-2\beta E}\rho_0(E)b\left(\frac{\tau}{2\pi \rho_0(E)}\right),
\label{eq:18}
\end{align}
where $b$ is the form factor and we have used notation consistent with \cite{Mehta2004}. The form factor depends on the specific symmetry class of the matrix model and we will use the value derived for the Gaussian orthogonal ensemble (GOE). To be consistent with the unorientable Airy model we again take the leading order energy density, $\rho_0(E)$,  to be given by~\eqref{eq:ed}. The exact result of the orthogonal SFF computed from universal RMT, referred to as $\kappa^{\text{GOE}}_{\beta}(\tau)$, is given in section~\ref{sec:SFF10}. Here we report the first few terms:
\begin{align}
    &{} e^{-S_0}\kappa_{\beta}^{\text{GOE}}(\tau) = \frac{\tau}{2\pi \beta}-\frac{\tau^2}{\sqrt{2\pi\beta}}-\frac{\gamma +\log\qty(2\beta \tau^2)+\frac{1}{3}}{\pi}\tau^3+\frac{8\sqrt{2\pi \beta}}{3\pi}\tau^4 +\order{\tau^5}.
    \label{eq:20}
\end{align}
In view of the definition \eqref{eq:18}, there cannot be any $t$-dependent terms.
\subsection{Comparison of unorientable topological gravity and universal RMT}
By comparison, the agreement of~\eqref{eq:19} and~\eqref{eq:20} is obvious for all terms except the $\tau^3$ coefficient. More generally, we have checked that all terms even in $\tau$ agree up to $\tau^8$, and all odd terms, in both computations, have a structure similar to the $\tau^3$ terms presented in~\eqref{eq:19} and~\eqref{eq:20}, respectively. For example, we have checked the coefficient of $\log(\beta)$ agrees up to $\tau^7$. In section \ref{sec:comparison} we prove the following equivalence for $\tau^3$:
\begin{align}
      - \tau^3\frac{\log \left(2\beta  \tau ^2\right)+\gamma+\frac{1}{3}}{\pi}=&  \frac{\tau^3}{\pi}\left(-\frac{10}{3}+\log \left(\frac{2t}{\beta}\right)-\frac{\sqrt{2\pi}}{3}\left(t\tau^2\right)^{1/2}  -\frac{2\left(t\tau^2\right)^2}{45} \right) \nonumber\\
      &+\order{t^{-1/2}} +\order{\tau^9},
\end{align}
where the terms of order $\tau^9$ and greater depend on $t$ and are not relevant for the $\tau^3$ comparison. The $\order{t^{-1/2}}$ indicates this equivalence is only valid in the limit $t \to \infty$. Therefore, the $\tau^3$ term also agrees in the universal limit. We further show that in order to prove the equivalence of the $\tau^5$ term, the $g=4$ from the loop equations, i.e. $V^{\text{Airy}}_{4,2}$, is needed, 
which, however, has not been computed so far. The agreement of terms even in $\tau$, the agreement of the $\tau^1$ and $\tau^3$ terms, and the structure of the rest of the odd terms provide strong evidence that the two results for the SFF agree in the universal limit.
\subsection{Cancellations}
We show directly that this agreement, as for the unitary case, implies the cancellation of certain contributions to the canonical SFF that would arise for a generic choice of the coefficients in the unorientable WP volumes. Along the lines of \cite{Weber2022,Blommaert2022}, we derive expressions for exemplary cases of said contributions and the corresponding constraints on the coefficients of the unorientable Airy WP volumes, which we show to be fulfilled in all the cases studied so far.

\subsection{Structure of the paper}
The rest of the paper is organized as follows: In the following section we briefly review the loop equations formalism, before demonstrating how the relevant contour integrals can be solved for the case of the unorientable Airy model. We then compute the unorientable Airy WP volumes, comment on their general structure, and use them to compute the canonical SFF. In section~\ref{sec:RMTUniversality} the general formula for the universal canonical SFF from RMT is derived and then computed for the specific case of the Airy model for the orthogonal symmetry class. In section~\ref{sec:comparison} we compare the two expressions, demonstrate equivalence for the first few terms in the series, and argue they are equivalent to all orders. Section \ref{sec:Cancellations} is concerned with giving an outlook on how to use the consistency of the unorientable Airy model with universal RMT to derive constraints on the coefficients of the unorientable Airy WP volumes. In \cref{app:Resolvents} and \cref{app:Volumes} we collect our results for the resolvents computed from the loop equations. In \cref{sec:Diagrammatics} we use diagrammatic considerations to show that the structure of the unorientable Airy WP volumes we claim is indeed generic. In \cref{sec:SFF_general} we work out the generic form of the canonical SFF for the unorientable Airy model as needed for the construction of constraints in \cref{sec:Cancellations}. Lastly, \cref{rmt_derive} contains the derivation of the universal RMT result for the canonical SFF of the unorientable Airy model.

\section{Canonical SFF from unorientable topological gravity }\label{sec:AiryWP}
\subsection{Matrix model computation}
As we explained in the introduction, it is possible to compute the correlation functions of partition functions of the unorientable Airy model by employing the duality to a double-scaled matrix model with the leading genus density of states $\rho^{\text{Airy}}_0(x) = \frac{\sqrt{x}}{2\pi}$\footnote{This genus $0$ density of states is the same for both the orientable and the unorientable theory.}. In order to do so, it is easiest to use the loop equations approach to compute the perturbative expansion of matrix model correlation functions of resolvents which then determine the unorientable Airy WP volumes needed for the computation of the correlation functions of partition functions. To facilitate presenting this computation, it is worthwhile to quickly recall said formalism for a not yet double-scaled one-cut matrix model, as this will be sufficient for the present case of interest. Readers already familiar with this formalism can skip this presentation and continue with \cref{sec:doubleCover}.

\subsubsection{The perturbative loop equations}
We follow the notation of \cite{Stanford2019}. There it was shown that for a matrix model of size $N$, determined by a potential $V(x)$ and a choice of $\upbeta\in\qty{1,2,4}$, where the choice of $\upbeta=1$ corresponds to the orthogonal case, it holds that 
\begin{equation}\label{eq:LoopEq}
			\begin{aligned}
			-N\ev{ P(x,I)}_c=&\left(1-\frac 2 \upbeta\right)\partial_x\ev{ R(x,I)}_c+\ev{ R(x,x,I)}_c+\sum_{J\supseteq I}\ev{ R(x,J)}_c \ev{R(x,I\backslash J)}_c\\
			& - N V^{\prime}(x)\ev{R(x,I)}_c+\frac 2 \upbeta \sum_{k=2}^{n}\partial_{x_k}\left[\frac{\ev{ R(x,I\backslash\left\{x_k\right\})}_c-\ev{ R(I)}_c}{x-x_k}\right],
			\end{aligned}
\end{equation}
where \begin{align}
    R(x_1,\ldots,x_n)=\Tr\frac{1}{x_1-H}\ldots\Tr\frac{1}{x_n-H}
\end{align} 
is the $n$-point resolvent and
\begin{align}
    I = \{x_1,\dots,x_n\}.
\end{align}
The definition of $P(J)$ can be looked up in \cite{Stanford2019}, but the important thing to note is that it is analytic in $x$. The idea now is to plug the genus expansion of the correlation functions of the resolvents, given by 
\begin{align}
    \expval{R(I)}_c=\sum_{g\in\qty{0,\frac 12 ,1, \dots,}} \frac{R_g(I)}{N^{2g+\abs{I}-2}},
\end{align}
into \cref{eq:LoopEq}. Considering each order in $N$ of the resulting equation separately enables one to compute $R_g(I)$ recursively through a dispersion-relation-like contour integral. For completeness, we swiftly recall the special cases $g=0$, $n=1,2,3$, computed in \cite{Stanford2019}, before citing the result for generic $g,n$.
\paragraph{$\mathbf{g=0,n=1}$}

Considering $I=\emptyset$ and taking the leading order term in \cref{eq:LoopEq}, one finds 
\begin{align}
    y(x)^2&=\left(\text{analytic}\right),\\
    \text{with} \ y(x):&=R_0(x)-\frac{V^\prime(x)}{2},\label{def:spectralCurve}
\end{align}
where $y(x)$ is the spectral curve which turns out to determine all the $R_g(I)$. The first special case encodes the connection of the spectral curve with the potential $V$, used here to define the matrix model. Furthermore, as the potential $V$ is assumed to be analytic, the relation can be used to replace subsequent occurrences of $R_0(x)$ by the spectral curve at the cost of adding an additional analytic term. For brevity's sake, we only state the properties of the spectral curve needed for the subsequent explanation, referring the reader for proof and details to e.g \cite{Stanford2019,Eynard2018}. As we are only concerned with one-cut matrix models, it suffices to consider the case of the cut being a real interval $\qty[a_-,a_+]$ with the spectral curve having a square-root singularity at the end points. Another important property of the spectral curve one can derive from the relation of $R_0(x)$ with the density of states is
\begin{align}
    \lim_{\epsilon\to 0} y(x\pm i \epsilon)=\mp i \pi \rho_0(x).
\end{align}
Hence, one can determine the spectral curve from $\rho_0$, i.e. the genus zero term, which for the case of interest in this paper is given by~\eqref{eq:ed}. This yields the spectral curve of the Airy model,
\begin{align}
    y^{\text{Airy}}(x)=\frac{\sqrt{-x}}{2},
\end{align}
i.e. $a_-=0,\ a_+\to \infty$.
\paragraph{$\mathbf{g=0,n=2}$}

\begin{figure}[h]
				\centering
				\begin{subfigure}[b]{0.35\textwidth}
					\includegraphics[width=\textwidth]{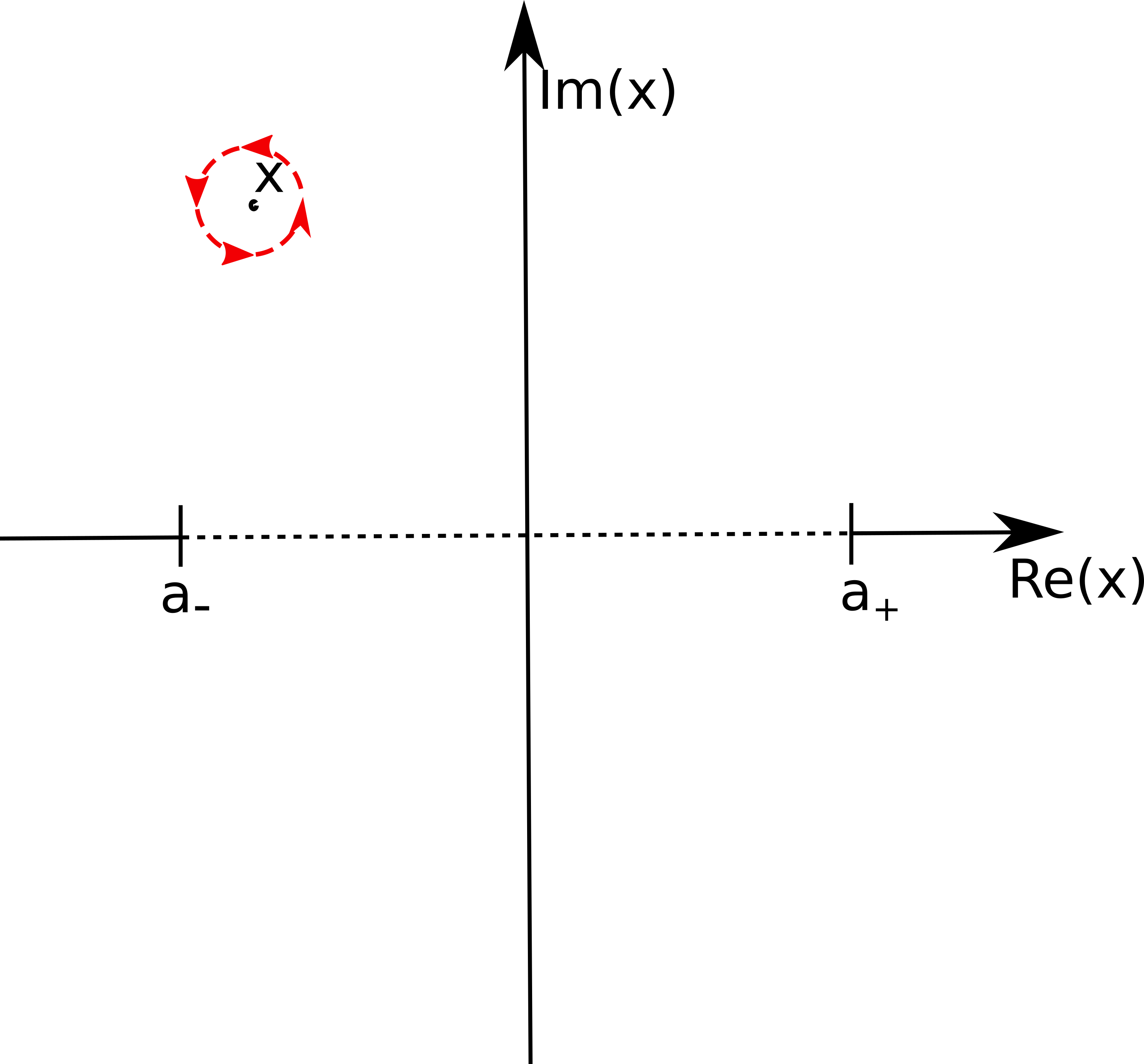}
					\caption{}
					\label{fig:Contour}
				\end{subfigure}
				\begin{subfigure}[b]{0.35\textwidth}
					\includegraphics[width=\textwidth]{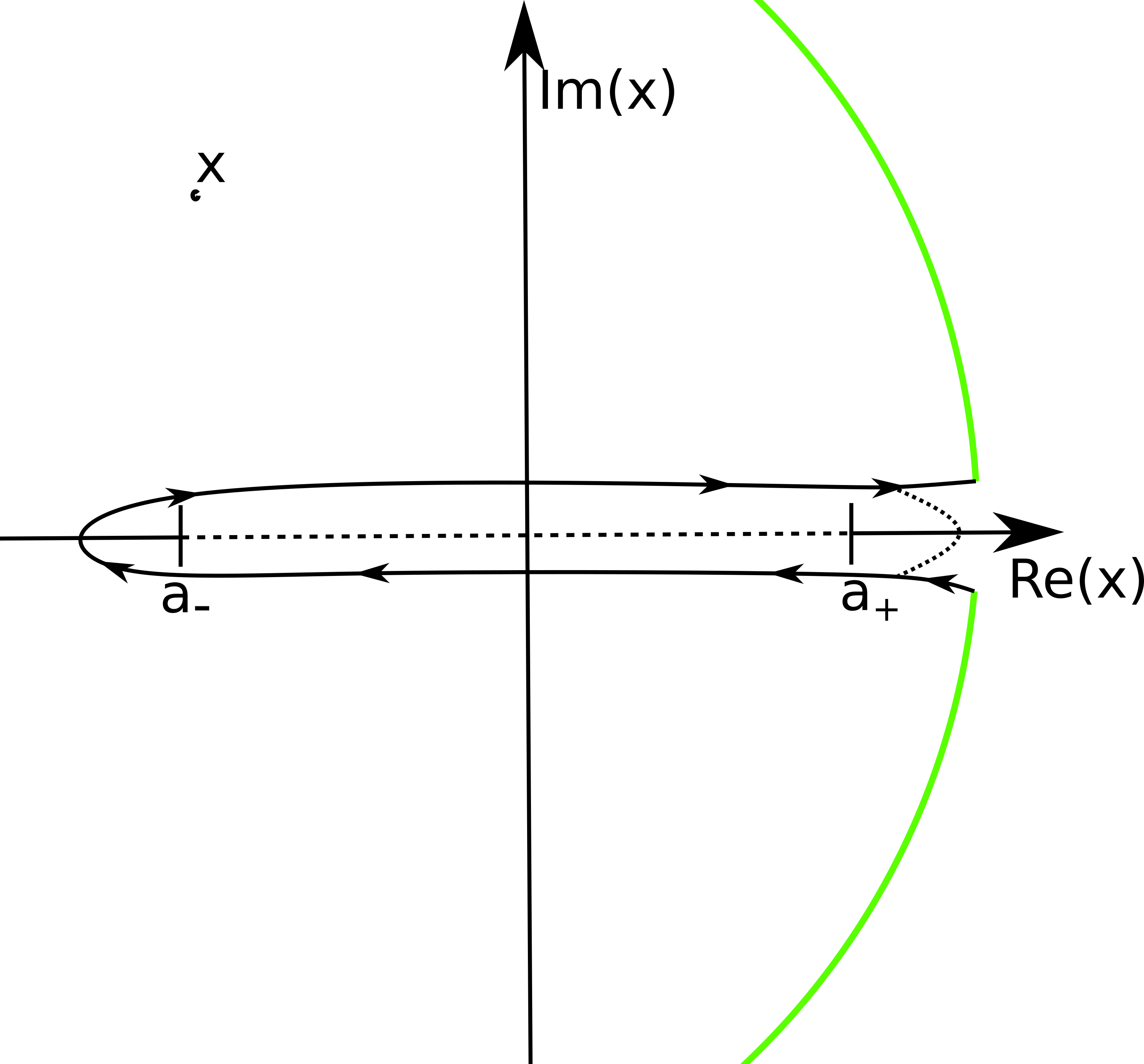}
					\caption{}
					\label{fig:ContourDef}
				\end{subfigure}
				\caption{Illustration of the branch cut structure of the spectral curve $y(x)$ of a one-cut matrix integral with the branch cut depicted as the broken line with the edges $a_-$ and $a_+$. In a) the contour $\mathcal{C}'$ is depicted in red, in b) its deformation is depicted in black with the final deformation $\mathcal{C}$ being the completion of the contour circumventing the cut, i.e. the black solid and dotted lines. }
				\label{fig:Contours}
			\end{figure}
Next, it is useful to consider $I=\qty{x_1}$ and look at the leading term in \cref{eq:LoopEq} arising for this case:
\begin{align}
    R_0(x,x_1)\underbrace{\left[2R_0(x)-V^\prime(x)\right]}_{=2y(x)}+\frac 2 \upbeta\partial_{x_1}\left[\frac{R_0(x)+R_0(x_1)}{x-x_1}\right]=\left(\text{analytic}\right).
\end{align}
Using \cref{def:spectralCurve} this can be rewritten as
\begin{align}
    2y(x)R_0(x,x_1)+\frac 2 \upbeta \frac{y(x)}{\left(x-x_1\right)^2}=\left(\text{analytic in}\ x\  \text{near the cut}\right).\label{eq:LE_02}
\end{align}
To solve for $R_0(x,x_1)$, we use an argument due to Migdal \cite{Migdal1983}.
The idea is to compute $R_0(x,x_1)$ by a suitable contour integral with an integrand vanishing at infinity. It is thus advisable to divide \cref{eq:LE_02} by $y(x)$, as in general, $y(x)$ does not vanish for large $\abs{x}$.
However, doing so alters the analyticity properties of the RHS, since $\frac{1}{y(x)}$ diverges at the edges of the cut. To cure this, one defines
\begin{align}
    \sigma(x)\coloneqq \qty(x-a_+)\qty(x-a_-),
\end{align}
such that $\sqrt{\sigma(x)}$ has the same branch-cut structure as $y(x)$\footnote{Meaning $\sqrt{\sigma(x)}=-\sqrt{\sigma(\widehat{x})}$, where $x$ and $\widehat x$ denote, as in \cite{Stanford2019}, the coordinate on the ``first'' and the ``second'' sheet, respectively.}, guaranteeing $\frac{\sqrt{\sigma(x)}}{y(x)}$ is analytic near the cut. Using this to divide out $y(x)$ results in 
\begin{align}
    \sqrt{\sigma(x)} R_0(x,x_1)+\frac 1 \upbeta \frac{\sqrt{\sigma(x)}}{\left(x-x_1\right)^2}=\left(\text{analytic in}\ x\  \text{near the cut}\right).\label{eq:LE_02_mod}
\end{align}
Now one can compute
\begin{align}
\begin{aligned}\label{eq:R_0_2_integral}
    R_0(x,x_1)\sqrt{\sigma(x)}\overset{\text{Cauchy}}{=}&\frac{1}{2\pi i}\oint_{\mathcal{C'}} \frac{\dd x'}{x'-x}R_0(x',x_1)\sqrt{\sigma(x')}\\
    =&\frac{1}{2\pi i}\oint_{\mathcal{C}} \frac{\dd x'}{x'-x}R_0(x',x_1)\sqrt{\sigma(x')}\\
    =&-\frac{1}{2\pi i}\frac 1 \upbeta \oint_{\mathcal{C}} \frac{\dd x'}{x'-x}\frac{\sqrt{\sigma(x')}}{\qty(x'-x_1)^2}\\
    \eqqcolon& -\frac{1}{\upbeta}\frac{1}{2\pi i} \oint_{\mathcal{C}}\dd{x'}f(x'),
\end{aligned}
\end{align}
where, in the first line Cauchy's theorem was used to rewrite $R_0(x,x_1)$ as an integral over the contour $\mathcal{C'}$ encircling $x$ clockwise, see \cref{fig:Contour}. In the second line the contour was deformed first to the black contour using the fact that $R_0(x',x_1)$ is holomorphic away from the cut, then the points where the green curve starts were brought together on the real line away from the cut, using again the analyticity of the integrand there. Now one takes the radius of the circle to infinity, using that for large $\abs{x'}$, $R_0(x',x_1)\propto \frac{1}{x'}$. Thus, only the contour $\mathcal{C}$ encircling the cut remains, as depicted in \cref{fig:ContourDef}. Employing once more the analyticity of the integrand, this contour is contracted so that it only encircles the cut-region. In the third line of \cref{eq:R_0_2_integral}, we use \cref{eq:LE_02_mod}, noting that the integral over a function analytic near the cut vanishes. 

\begin{figure}[h]
		\centering
		\begin{subfigure}[b]{0.35\textwidth}
			\includegraphics[width=\textwidth]{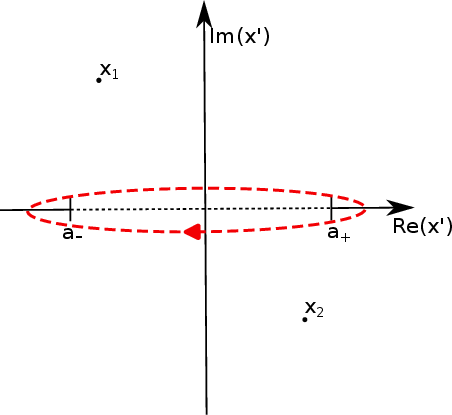}
			\caption{}
			\label{fig:Contour_R02}
		\end{subfigure}
		\begin{subfigure}[b]{0.35\textwidth}
			\includegraphics[width=\textwidth]{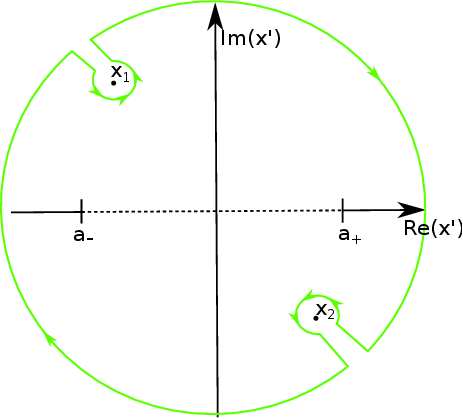}
			\caption{}
			\label{fig:ContourDef_R02}
		\end{subfigure}
		\caption{Illustration of the deformation of the integration contour. In a), the original contour $\mathcal{C}$ is depicted in red, in b), its deformation $\mathcal{C'}$ is depicted in green.}
		\label{fig:Contours_mod_R02}
	\end{figure}

To evaluate this integral, it is most convenient to deform the contour $\mathcal{C}$ to the contour $\mathcal{C'}$, as depicted in \cref{fig:Contours_mod_R02}.
The deformed contour $\mathcal{C'}$ is chosen such that the contribution of the circle vanishes as its radius is sent to infinity due to the behaviour of the integrand for large $x'$. Taking the connections from the arc at infinity to the small circles around $x_1$ and $x_2$ very narrow, the two parts of a given connection will cancel each other (as they run in opposite directions and the integrand is analytic in the region enclosed by them). Hence, one is left only with the contribution of the two poles of the integrand at $x_1$ and $x_2$, which are encircled in the mathematically positive way. Using the residue theorem, one finds
\begin{align}
    \begin{aligned}
        R_0(x_1,x_2)\sqrt{\sigma(x_1)}&=-\frac{1}{\upbeta}\qty(\underset{x'=x_1}{\Res}+\underset{x'=x_2}{\Res})f(x')\\
	&=-\frac{1}{\upbeta}\qty[\frac{\sqrt{\sigma(x_1)}}{\qty(x_1-x_2)^2}-\frac{x_1x_2+a_+ a_- -\frac{a_+ +a_- }{2}\qty(x_1+x_2)}{\qty(x_1-x_2)^2\sqrt{\sigma(x_2)}}]\\
	&=\frac{1}{\upbeta \qty(x_1-x_2)^2}\qty[\frac{x_1x_2+a_+ a_- -\frac{a_+ +a_- }{2}\qty(x_1+x_2)}{\sqrt{\sigma(x_2)}}-\sqrt{\sigma(x_1)}].
     \end{aligned}
\end{align}
Hence, one has
\begin{align}
    R_0(x_1,x_2)=\frac{1}{\upbeta\qty(x_1-x_2)^2}\qty(\frac{x_1x_2-\frac{a_+ + a_-}{2}\qty(x_1+x_2)+a_+a_-}{\sqrt{\sigma(x_1)}\sqrt{\sigma(x_2)}}-1), \label{eq:R_0_2_N}
\end{align}
which only depends on the edges of the cut that are specified when giving a spectral curve. For the spectral curve  $y^{\text{Airy}}(x)$, one has to consider the limits $a_-\rightarrow 0$, $a_+\rightarrow \infty$. Taking these limits one after the other in \cref{eq:R_0_2_N}, one finds
\begin{equation}
\begin{aligned}
    R_0(x_1,x_2)&=\frac{1}{\upbeta\qty(x_1-x_2)^2}\qty(\frac{x_1x_2-\frac{a_+}{2}\qty(x_1+x_2)}{\sqrt{x_1^2-x_1 a_+}\sqrt{x_2^2-x_2a_+}}-1)\\
    &\xrightarrow{a_+\to\infty
    }\frac{1}{2\upbeta\qty[\qty(\sqrt{-x_1}-\sqrt{x_2})\qty(\sqrt{-x_1}+\sqrt{-x_2})]^2}\qty(\frac{-x_1-x_2-2\sqrt{-x_1}\sqrt{-x_2}}{\sqrt{-x_1}\sqrt{-x_2}})\\
    &=\frac{1}{2\upbeta}\frac{1}{\qty(\sqrt{-x_1}+\sqrt{-x_2})^2\sqrt{-x_1}\sqrt{-x_2}},\label{eq:R_02RMT}
\end{aligned}
\end{equation}
as reported in \cite{Stanford2019}.
\paragraph{$\mathbf{g=0,n=3}$}
For the last special case, $\qty{x_1,x_2}$, analogously to the previous case, one considers the leading term and rewrites it as a contour integral to find
\begin{align}
\begin{aligned}
    R_0\qty(x,x_1,x_2)=&-\frac{1}{2\pi i}\int_\mathcal{C}\frac{\dd{x'}}{x'-x}\frac{\sqrt{\sigma(x')}}{\sqrt{\sigma\qty(x)}}\frac{1}{y(x')}\times\\
    &\qty[R_0(x',x_1)R_0(x',x_2)+\frac{1}{\upbeta}\qty(\frac{R_0(x',x_2)}{\qty(x'-x_1)^2}+\frac{R_0(x',x_1)}{\qty(x'-x_2)^2})].\label{eq:R_0_3_N}
\end{aligned}
\end{align}

\paragraph{\textbf{Generic case}}
For all other cases, one can show in analogous fashion that
\begin{align}
    R_g(x,I)=-\frac{1}{2\pi i}\oint_{\mathcal{C}}\frac{\dd x'}{x'-x}\frac{\sqrt{\sigma(x')}}{\sqrt{\sigma\qty(x)}}\frac{1}{2y(x')}F_g(x',I),\label{eq:RecursionRInt}
\end{align}
with
\begin{align}
    \begin{aligned}
			F_g(x,I)=&\qty(1-\frac 2 \upbeta)\partial_x R_{g-\frac 12}(x,I)+R_{g-1}(x,x,I)\\
					&+\sum'_{J\supseteq I, h} R_h(x,J)R_{g-h}(x,I\backslash J)\\
					&+2\sum_{k=1}^{n}\qty[R_0(x,x_k)+\frac 1 \upbeta \frac{1}{\qty(x-x_k)^2}]R_{g}(x,I\backslash\qty{x_k}),\label{eq:F_g(x)}
			\end{aligned}
\end{align}
where $\sum'$ indicates that $R_0(x)$ and $R_0(x,x_k)$ are excluded from the sum\footnote{Note that this formula needs to be modified for $g=0$, hence why we split off the special cases $(g,n)$=$(0,1), \ (0,2),\ (0,3)$. We do not provide them, as for higher $n $ it is more efficient to use the quite fast algorithm of \cite{Do2009} for the orientable volumes and translate to the unorientable volumes by the simple relation derived in \cite{Stanford2019}.}. 

\subsubsection{Introduction of the double cover coordinate}\label{sec:doubleCover}
To solve the contour integrals, as demonstrated already in the $g=0,n=2$ case, one could deform the contour $\mathcal{C}$ to surround the poles of the integrand, enabling evaluation by the residue theorem. However, it turns out to be more convenient to use the double-cover coordinate $z\in \mathbb{P}^1$,\footnote{Using the notation $\mathbb P ^1$ for the Riemann sphere}(e.g. \cite{Post2022}), defined via $x=-z^2$, to rewrite the integrals, which will be the object of this section.

Before doing so, we briefly comment on the fate of the $\sqrt{\sigma}$ terms in the case $a_+\to \infty, \ a_-= 0$. As seen from~\eqref{eq:RecursionRInt}, we only have to consider the term,
\begin{align}
   \lim\limits_{a_+\to\infty} \frac{\sqrt{x'\qty(x'-a_+)}}{\sqrt{x\qty(x-a_+)}}=\frac{\sqrt{-x'}}{ \sqrt{-x}}=\frac{z'}{z},
\end{align}
where we went to the double-cover coordinate in the last step. Furthermore, the spectral curve becomes $y^{\text{Airy}}(z)=\frac 12 z$.

From here, it is straightforward to write down \cref{eq:RecursionRInt} in the double cover coordinates
\begin{align}
\begin{aligned}\label{eq:RecursionRInt_z}
    R_g(-z^2,I)&=-\frac{1}{2\pi i}\oint_{\mathcal{C}_z}\frac{2z'\dd z'}{{z'}^2-z^2}\frac{z'}{z}\frac{1}{2y(-z'^2)}F_g(-z'^2,I)\\
    &=-\frac{1}{2\pi i z}\oint_{\mathcal{C}_z}\frac{{z'}^2\dd z'}{{z'}^2-z^2}\frac{1}{y(-{z'}^2)}F_g(-z'^2,I)\\
    &=\frac{1}{2\pi i z}\oint_{\qty[-i \infty+\epsilon,i\infty+\epsilon]}\frac{{z'}^2\dd z'}{{z'}^2-z^2}\frac{1}{y(-{z'}^2)}F_g(-z'^2,I)
\end{aligned}
\end{align}
Here, $\mathcal{C}_z$ is the preimage of $\mathcal{C}$ under $z\to x(z)$. In the last line this has been made explicit as the interval $\qty[-i \infty+\epsilon,i\infty+\epsilon]$ which, as $z$ is defined on the Riemann sphere $\mathbb{P}^1$, is indeed a closed contour winding once around the sphere with a small offset to the hemisphere of positive real value. One can argue that this is indeed the correct preimage  as follows. The most important property of the contour $\mathcal{C}$ is that it encircles the cut once in a clock-wise fashion. Thus, for $a_- =0$, $a_+\to \infty$, it intersects the negative real axis at a position $-\delta$ and before (after) this intersection point has negative (positive) imaginary part. Using the branch-cut structure of the square root, one can quickly convince oneself that the contour in the double cover coordinate has to go from positive imaginary infinity to negative imaginary infinity with a small offset $\epsilon$ which gives the claimed contour. Changing the direction of integration, thus cancelling the sign, gives the result stated above, which is in agreement with \cite{Stanford2023}. In the following we will denote this contour in an abbreviated fashion as $i\mathbb{R}+ \epsilon$.

Now consider $F_g(x,I)$. After noting that 
\begin{align}
    \partial_x f(x)=\frac{1}{-2z}\partial_z f(z),
\end{align}
one finds
\begin{align}\label{eq:F_g(z)}
    \begin{aligned}
			F_g(-z^2,I)=&\qty(1-\frac 2 \upbeta)\frac{1}{-2z}\partial_z R_{g-\frac 12}(-z^2,I)+R_{g-1}(-z^2,-z^2,I)\\
					&+\sum'_{I\supseteq J, h} R_h(-z^2,J)R_{g-h}(-z^2,I\backslash J)\\
					&+2\sum_{k=1}^{n}\qty[R_0(-z^2,-z^2_k)+\frac 1 \upbeta \frac{1}{\qty(z_k^2-z^2)^2}]R_{g}(-z^2,I\backslash\qty{-z^2_k}).
			\end{aligned}
\end{align}
For the sake of readability, we will denote the dependence on a double-cover coordinate $z_i$ just by $z_i$, making the true dependence on $-z_i^2$ implicit from now on.

Having considered the general case of $F_g(z,I)$ in the double-cover coordinate system, it is illustrative to reconsider the case of $(g,n)=(0,2)$. Transforming the relevant contour integral for this case, given in \cref{eq:R_0_2_integral}, to the double-cover coordinate one finds
\begin{align}
    R_0(z_1, z_2) =\frac{1}{2\pi i \upbeta z_1} \oint_{i\mathbb{R}+\epsilon}\frac{\dd{z'}}{z'^2-z_1^2}\frac{2 z'^2}{\qty(z'^2-z_2^2)^2}
\end{align}
The integrand as a function of $z'$ has poles at $\pm z_1$ and $\pm z_2$. We first consider the case of $\Re(z_1)>0$ and $\Re(z_2)>0$\footnote{Actually larger than $\epsilon$, but by sending $\epsilon \to 0$ after the computation this drops out.}. Then, the poles at $-z_1$ and $-z_2$ are in the hemisphere of the negative real part (left) while the other two are on the other hemisphere. Now there are two ways to deform the contour, first surrounding the two poles on the left hemisphere counterclockwise, second surrounding those on the other hemisphere clockwise. The first option gives
\begin{align}
\begin{aligned}
    R_0(z_1, z_2)&=\frac{1}{\upbeta z_1}\qty[\underset{z'=-z_1}{\Res}+\underset{z'=-z_2}{\Res}]\frac{\dd{z'}}{z'^2-z_1^2}\frac{2 z'^2}{\qty(z'^2-z_2^2)^2}\\
    &=\frac{1}{2\upbeta}\frac{1}{z_1 z_2 \qty(z_1+z_2)^2},
\end{aligned}
\end{align}
and the second
\begin{align}
\begin{aligned}
    R_0(z_1, z_2)&=\frac{-1}{\upbeta z_1}\qty[\underset{z'=z_1}{\Res}+\underset{z'=z_2}{\Res}]\frac{\dd{z'}}{z'^2-z_1^2}\frac{2 z'^2}{\qty(z'^2-z_2^2)^2}\\
    &=\frac{1}{2\upbeta}\frac{1}{z_1 z_2 \qty(z_1+z_2)^2},\label{eq:R_02_z}
\end{aligned}
\end{align}
which agree with each other, as they must, and with~\eqref{eq:R_0_2_N}. The choice of positive real parts of the double-cover coordinate yields $R_{0,2}$ on the ``first sheet'' while other choices yield the different continuations to other sheets. We are only interested in the ``first sheet'' quantities, so we will use the assumption $\Re(z_i)>0$ to compute the resolvents and then extend the result to the whole plane in each variable.

This concludes the discussion of the loop equations for a general one-cut double-scaled matrix model. The remainder of this section will focus on applying the formalism for the Airy spectral curve in the orthogonal symmetry class.
\subsubsection{Airy model correlation functions for $\upbeta=1$}
Before computing the correlation functions of resolvents, it is worthwhile to quickly recall their relation with the Airy WP volumes \footnote{This relation holds for the orientable as well as the unorientable case.} to facilitate the comparison to the few known results in the literature. The relation of the (Airy) WP volumes with the contributions to the genus expansion of resolvents is given by
\begin{align}
    V_{g,n}(b_1,\dots,b_n)&=\mathcal{L}^{-1}\qty[\prod_{i=1}^n\qty( \frac{-2z_i}{b_i})R_g(-z_1^2,\dots,-z_n^2),\qty(b_1,\dots,b_n)] \\
    &=(-1)^n\int_{\delta +i\mathbb{R}}R_g\left(-z_1^2,\dots,-z_n^2\right)\prod_{j=1}^n \frac{dz_j}{2\pi i}\frac{2z_j}{b_j}e^{b_j z_j} 
    \label{eq:V[R]}
\end{align}
as one can quickly derive by inverting the expression of the partition function and inserting the JT gravity formula for the partition function as trumpets integrated against the relevant WP volume. However, the case of $g=\frac 12,n=1$ is special, as one has to integrate the trumpet only against $\dd b$ (as opposed to the usual WP form $b\,\dd b$) to obtain the partition function \cite{Stanford2019}. Hence, in this case the above formula must be modified by removing the division by $b_1$.

Having stated this expression we begin by considering some examples for the computation of the resolvents in the orthogonal case, i.e. $\upbeta=1$.
\paragraph{$\mathbf{g=\frac12, n=1}$}~\\
In this arguably simplest case, one has
\begin{align}
    F_{\frac12}\qty(-z^2)=\frac{1}{2z}\partial_z R_0(-z^2)\rightarrow\frac{1}{2z}\partial_z y(z)=\frac{1}{4 z},
\end{align}
where $R_0(-z^2)$ was replaced by the spectral curve by adding analytic terms which vanish under the contour integration. Thus by \cref{eq:RecursionRInt_z}
\begin{align}
    \begin{aligned}
        R_{\frac 12}\qty(z)&=\frac{1}{2\pi i z}\int_{i\mathbb{R}+\epsilon}\frac{{z'}^2 \dd{z'}}{{z'}^2-{z}^2}\frac{2}{z'}\frac{1}{4z'}=\frac 12 \frac{1}{2\pi i z}\int_{i\mathbb{R}+\epsilon}\frac{\dd{z'}}{\qty(z'-z)\qty(z+z')}\\
        & =\frac{1}{2 z}\underset{z'=-z}{\Res}\frac{1}{\qty(z'-z)\qty(z'+z)}=-\frac{1}{2 z}\underset{z'=z}{\Res}\frac{1}{\qty(z'-z)\qty(z'+z)}\\
        &=-\frac{1}{4z^2},
    \end{aligned}
\end{align}
where in the evaluation of the contour integral, the assumption $\Re{z}>0$ was used.
To compare with the JT result from \cite{Stanford2019}, we compute the corresponding unorientable Airy WP volume, i.e.
\begin{align}
    V^{\text{Airy}}_\frac 12 (b)=\mathcal{L}^{-1}\qty[-2z R_{\frac 12}(z),b]=\frac 12 \mathcal{L}^{-1}\qty[\frac 1 z,b]=\frac 12.
\end{align}
In the JT case one finds \cite{Stanford2019} $V_\frac 12 (b)=\frac 12 \coth{\frac b 4}$, which, in the large $b$ limit, reproduces the unorientable Airy result\footnote{This is not surprising, since that result could be derived using the same integral as here, only using the JT spectral curve.}.

\paragraph{$\mathbf{g=1,n=1}$}~\\
Here one finds from \cref{eq:F_g(z)}
\begin{align}
    F_{1}(z')=\frac{\partial_{z'} R_{\frac{1}{2}}(z')}{2 z'}+R_{0,2}(z',z')+R_{\frac{1}{2}}(z') R_{\frac{1}{2}}(z')=\cdots=\frac{7}{16 {z'}^4}
\end{align}
and thus
\begin{align}
\begin{aligned}
    R_{1}\qty(z)&=\frac{1}{2\pi i z}\int_{i\mathbb{R}+\epsilon}\frac{{z'}^2 \dd{z'}}{{z'}^2-{z}^2}\frac{2}{z'}\frac{7}{16 {z'}^4}=\frac{7}{8z}\frac{1}{2\pi i}\int_{i\mathbb{R}+\epsilon}\frac{ \dd{z'}}{\qty(z'-z)\qty(z'+z){z'}^3}\\
    &=\frac{7}{8z}\qty[\underset{z'=0}{\Res}+\underset{z'=-z}{\Res}]\frac{ \dd{z'}}{\qty(z'-z)\qty(z'+z){z'}^3}=\frac{7}{8z}\qty(\frac{1}{2{z}^4}-\frac{1}{z^4})=-\frac{7}{16 z^5}\\
    &=-\frac{7}{8z}\underset{z'=z}{\Res}\frac{ \dd{z'}}{\qty(z'-z)\qty(z'+z){z'}^3}=-\frac{7}{16 z^5}.
\end{aligned}
\end{align}
With this, one can compute the corresponding WP volume
\begin{align}
    V^{\text{Airy}}_1 (b)=\frac{1}{b} \mathcal{L}^{-1}\qty[-2z R_1(z),b]=\frac{7}{8b}\mathcal{L}^{-1}\qty[z^{-4},b]=\frac{7}{8 \cdot 3!}b^2=\frac{7}{48}b^2,
\end{align}
which matches the result reported in \cite{Stanford2023}.
\paragraph{$\mathbf{g=\frac{1}{2} , n=2}$}~\\
For this case, one has
\begin{align}
    F_\frac 12 (z',z_2)=\frac{1}{2z'}\partial_{z'}R_0(z',z_2)+2 R_\frac 12 (z')\qty[R_0(z',z_2)+\frac{1}{\qty({z'}^2-z_2^2)^2}],
\end{align}
which, using \cref{eq:RecursionRInt_z}, leads to
\begin{align}
\begin{aligned}
    R_{\frac 12}(z_1,z_2)&=\frac{1}{z_1}\qty[\underset{z'=0}{\Res}+\underset{z'=-z_1}{\Res}+\underset{z'=-z_2}{\Res}](-1)\frac{z_2 \left(z_2 \left(z_2+z'\right)-2 z'^2\right)+2 z'^3}{ z_2 z'^2 \left(z'^2-z_1^2\right) \left(z'-z_2\right){}^2 \left(z_2+z'\right){}^3}\\
    &=-\frac{1}{z_1}\qty[\underset{z'=z_1}{\Res}+\underset{z'=z_2}{\Res}](-1)\frac{z_2 \left(z_2 \left(z_2+z'\right)-2 z'^2\right)+2 z'^3}{ z_2 z'^2 \left(z'^2-z_1^2\right) \left(z'-z_2\right){}^2 \left(z_2+z'\right){}^3}\\
    &=\frac{z_1^4+3 z_2 z_1^3+3 z_2^2 z_1^2+3 z_2^3 z_1+z_2^4}{2 z_1^4 z_2^4 \left(z_1+z_2\right){}^3}.
\end{aligned}
\end{align}
Again using \cref{eq:V[R]} to compute the WP volume, one finds
\begin{align}
    V^{\text{Airy}}_\frac 12(b_1,b_2)=\theta\qty(b_1-b_2)b_1+\theta\qty(b_2-b_1)b_2=\max\qty(b_1,b_2),\label{eq:V_12_2}
\end{align}
in accordance with the result reported in \cite{Saad2022}.
\paragraph{$\mathbf{g=0,n=3}$}~\\
Remembering that this was a special case of the loop equations, we first rewrite \cref{eq:R_0_3_N} in the double-cover coordinate to find
\begin{align}
    \begin{aligned}
        R_0\qty(z_1,z_2,z_3)&=\frac{1}{2\pi i z_1}\oint_{i\mathbb R +\epsilon}\frac{z'^2\dd z'}{{z'}^2-z_1^2}\frac{2}{y(z')}\\
        &\qty[R_0(z',z_2)R_0(z',z_3)+\frac{1}{\upbeta}\qty(\frac{R_0(z',z_3)}{\qty(z_2^2-{z'}^2)^2}+\frac{R_0(z',z_2)}{\qty(z_3^2-{z'}^2)^2})] \label{eq:R_03_z}.
    \end{aligned}
\end{align}
This can now be evaluated as usual to yield
\begin{align}
    \begin{aligned}
        R_0\qty(z_1,z_2,z_3)&=\frac{1}{z_1} \qty[\underset{z'=0}{\Res}+\underset{z'=-z_1}{\Res}+\underset{z'=-z_2}{\Res}+\underset{z'=-z_3}{\Res}]\frac{z'^2}{{z'}^2-z_1^2}\frac{2}{y(z')}\\
        &\qty[R_0(z',z_2)R_0(z',z_3)+\qty(\frac{R_0(z',z_3)}{\qty(z_2^2-{z'}^2)^2}+\frac{R_0(z',z_2)}{\qty(z_3^2-{z'}^2)^2})]\\
        &=\frac{1}{z_1} \qty[\underset{z'=0}{\Res}+\underset{z'=-z_1}{\Res}+\underset{z'=-z_2}{\Res}+\underset{z'=-z_3}{\Res}]\frac{z_3^2 z'^2-4 z_2 z_3 z'^2+z_2^2 \left(z_3^2+z'^2\right)+z'^4}{ z_2 z_3 z' \left(z'^2-z_1^2\right) \left(z'^2-z_2^2\right){}^2 \left(z'^2-z_3^2\right){}^2}\\
        &=-\frac{1}{2 z_1^3 z_2^3 z_3^3} .
    \end{aligned}
\end{align}
Of course, one finds the same result when using the deformation of the contour surrounding the poles at $z_i$ in the right hemisphere. For the corresponding WP volume, we have
\begin{align}
    V^{\text{Airy}}_{0,3}(b_1,b_2,b_3)=4=2^{3-1}V^{\text{Airy, }\upbeta=2}_{0,3},
\end{align}
which matches the rule, proven in \cite{Stanford2019}, that for $g=0$, 
\begin{align}
V_0^{\upbeta=1}(I)=2^{\abs{I}-1}V_0^{\upbeta=2}(I).
\end{align}

These examples should be sufficient to illustrate the method used to compute the resolvents and, from those, the unorientable Airy WP volumes. As the main object of interest of this paper is the spectral form factor, we report the resolvents for $n=2$ in \cref{app:Resolvents}. To find the ones appearing in the recursion, we refer to the Mathematica notebook in the supplementary material. Rather than reporting their lengthy expressions, here we give the general form of the resolvents which, at fixed genus $g$, is given by
\begin{align}
    R_g(z_1,z_2)=\frac{P_g(z_1,z_2)}{(z_1 z_2)^{6g+1} (z_1+z_2)^{2g+2}},\label{eq:R_g_general}
\end{align}
with a symmetric polynomial $P_g(z_1,z_2)$ of combined degree $8g$. This form can be read off from the results of the computation and can also be motivated by the diagrammatic considerations in \cref{sec:Diagrammatics}. \footnote{To be more specific, the diagrammatics provide one with a proof for the form, while the orders of the polynomial $P_g(z_1,z_2)$ and the order in $(z_1+z_2)$ is predicted higher than observed from the result of the loop equations. This however is not a contradiction, but rather implies that the polynomial appearing in the numerator of the diagrammatic result must have as many zeroes at $z_1=-z_2$ as factors of $(z_1+z_2)$ required to cancel in order to be consistent with the loop equations. This can of course only be achieved if the coefficients of the diagrams have a certain type of symmetry, which is further explored in \cref{sec:Diagrammatics}.}

In the following section, we use the results of the resolvents from the loop equations to compute the unorientable Airy WP volumes for $n=2$ up to $g=\frac 7 2$.
\subsection{Unorientable Airy WP volumes}
Before reporting the volumes, let us first make some general statements about their form.

In the unitary case, i.e. for orientable manifolds, the Airy WP volumes for two geodesic boundaries have the form \cite{Mirzakhani2006}
\begin{align}
	V^{\text{Airy},\upbeta=2}_{g,2}(b_1,b_2)=\sum_{\alpha_1,\alpha_2\in \mathbb{N}_0}^{\alpha_1+\alpha_2=3g-1}C^{\upbeta=2}_{\alpha_1,\alpha_2}b_1^{2\alpha_1}b_2^{2\alpha_2},\label{eq:V_orientable}
\end{align}
with $C^{\upbeta=2}_{\alpha_1,\alpha_2}\in \mathbb{Q}_+$. The immediate generalisation for a larger number of boundaries also holds true, but we focus here on $n=2$ as a sufficient example. Furthermore, the coefficients $C^{\upbeta=2}_{\alpha_1,\alpha_2}$ are symmetric. The question at hand is how this is modified for unorientable manifolds.

The mathematical foundation of the full WP volumes in the unorientable case is less established than in the orientable case, i.e. their computation requires a regularization. The unorientable Airy WP volumes, in which we are interested here, do not suffer from this as they are precisely the first regularization independent terms of the regularized unorientable WP volumes \cite{Stanford2023}. However, to the best of our knowledge, the general form of the unorientable Airy WP volumes (i.e. an expression like \cref{eq:V_orientable}) was not yet given in the literature. Working towards this, we begin by studying which properties of the orientable Airy WP volumes should carry over to the unorientable setting.
First of all, already by the definition of the correlation functions, the general form of the resolvents reported in \cref{eq:R_g_general}, or the diagrammatic considerations of \cref{sec:Diagrammatics}, the symmetry under $b_1\leftrightarrow b_2$ is still present. Furthermore, one can argue, again using \cref{eq:R_g_general} or the diagrammatic expansion discussed in \cref{sec:Diagrammatics}, that for a fixed value of the lengths, the volumes are still polynomials. In general, those polynomials are not symmetric and thus the symmetry in $b_1\leftrightarrow b_2$ can only be retained if one introduces step functions that differentiate the cases $b_1>b_2$ and $b_1<b_2$. Therefore, we propose as a generalisation of \cref{eq:V_orientable}:
\begin{align}\label{eq:AiryWP_form_conjecture}
    V^{\text{Airy}(,\upbeta=1)}_{g,2}(b_1,b_2)=V^>_g(b_1,b_2)\theta\qty(b_1-b_2)+V^>_g(b_2,b_1)\theta\qty(b_2-b_1),
\end{align}
with 
\begin{align}
	V^{>}_{g}(b_1,b_2)=\sum_{\alpha_1,\alpha_2 \in \mathbb{N}_0}^{\alpha_1+\alpha_2= 6g-2}C_{\alpha_1,\alpha_2}b_1^{\alpha_1}b_2^{\alpha_2},
\end{align}
where the coefficients $C_{\alpha_1,\alpha_2}$ are not necessarily symmetric but still non-negative rational numbers\footnote{A possible source of confusion here is that the indices in $C_{\alpha_1,\alpha_2}$ correspond to the actual powers of the respective length in the respective monomial whereas in the orientable case the indices of $C^{\upbeta=2}_{\alpha_1,\alpha_2}$ correspond to half the powers of lengths in the respective monomial of the orientable Airy WP volume. This is used to be consistent with the literature convention for the orientable Airy WP volumes where this way of writing the coefficients makes more sense as only even orders of lengths appear in the volumes.}.
This is indeed reproduced by transforming the resolvents obtained in the previous chapter. Furthermore, in \cref{sec:Diagrammatics} diagrammatics are used to show that this is generic.

As specific examples we give the results for the lowest three non-zero genera while referring the reader to \cref{app:Volumes} for a complete list of the computed volumes.
\begin{align}
    V^{>}_{\frac 12}(b_1,b_2)&=b_1,\\
    V^{>}_{1}(b_1,b_2)&=\frac{1}{96} \qty(7 b_1^4+14 b_2^2 b_1^2+8 b_2^3 b_1+3 b_2^4 ),\\
    V^{>}_{\frac 32}(b_1,b_2)&=\frac{1}{40320}\qty(64 b_1^7+448 b_2^2 b_1^5+245 b_2^3 b_1^4+560 b_2^4 b_1^3+147 b_2^5 b_1^2+175 b_2^6 b_1+23 b_2^7).
\end{align}
In the next section, we will use the volumes found above to compute the spectral form factor.
\subsection{The canonical SFF at large times}\label{sec:SFF}
The canonical SFF is defined as 
\begin{align}
    \kappa_\beta(t)\coloneqq \ev{Z(\beta+i t)Z(\beta-i t)}_c
\end{align}
where $\beta\in \mathbb{R_+}$ is the inverse temperature and $t$ the time. In the following we will, for readability's sake, use $\beta_1=\beta+it$, $\beta_2=\beta_1^*$. The genus expansion of the correlation function of partition functions translates to a genus expansion of the spectral form factor, i.e. 
\begin{align}
    \kappa_\beta (t)=\sum_{g=0,\frac 12,1,\dots}e^{-2g S_0}Z_{g,2}(\beta_1,\beta_2)=\sum_{g=0,\frac 12,1,\dots}e^{-2g S_0}\kappa_\beta^g(t).
\end{align}
For a given genus $g$, the contribution to the correlation function of partition functions is given by
\begin{align}
    Z_{g,2}(\beta_1,\beta_2)=\int_0^\infty \dd{b_1}b_1 \int_0^\infty \dd{b_2}b_2 Z^t(\beta_1,b_1) Z^t(\beta_2,b_2) V^{\text{Airy}}_{g,2}(b_1,b_2)
\end{align}
with
\begin{align}
    Z^t(\beta,b)=\frac{1}{\sqrt{4\pi \beta}}e^{-\frac{b^2}{4\beta}}.
\end{align} 
Using this, one can compute the $\kappa_\beta^g(t)$ from the volumes stated above.
However, as we have already explained, the main point of interest of this work is to investigate the ``universal'' part of the spectral form factor which for the canonical SFF means the behaviour at large times, i.e. times of the order $e^{S_0}$. Thus, due to the prefactor $e^{-2g S_0}$ of the contribution to the spectral form factor at genus $g$, terms of a smaller or equal order in $t$ than $2g$ can be neglected. The relevant contributions (neglecting the lower order terms) to $\kappa_\beta(t)$ are given by
\begin{align}
    &\kappa^{0}_{\beta}(t)= \frac{\sqrt{t^2+\beta^2}}{2\pi \beta} \label{eq:k0}\\
    &\kappa^{\frac 12}_{\beta}(t)=-\frac{t^2+\beta^2}{\sqrt{2\pi \beta }}\label{eq:k1/2}\\
    &\kappa^{1}_{\beta}(t)=\qty[\frac{-10}{3}+i \qty(\arctan\qty(\sqrt{\frac{\beta- i t}{\beta + it}})-\arctan\qty(\sqrt{\frac{\beta + i t}{\beta - it}}))]\frac{t^3}{\pi}\\
    &\kappa^{\frac 3 2}_{\beta}(t)=\frac{8\sqrt{2\pi \beta}}{3\pi} t^4-\frac{i t^4}{3\sqrt{\pi}}\qty(\sqrt{\beta-i t}-\sqrt{\beta + it})\\
    &\kappa^{2}_{\beta}(t)=\frac{\beta t^5}{\pi}\qty[\frac{163}{15}-2\pi i +8i\arctan\qty(\sqrt{\frac{\beta + i t}{\beta - it}})]\\
    &\begin{aligned}
        \kappa^{\frac 5 2}_{\beta}(t)=&-\frac{64\qty(2\pi\beta)^{\frac 3 2}}{15\pi^2} t^6+\frac{t^6 \sqrt{t^2+\beta^2}}{30\sqrt{\pi}}\qty(\sqrt{\beta-i t}+\sqrt{\beta + it})\\
    &+\frac{ 21 i t^5 \sqrt{t^2+\beta^2}\beta}{5\sqrt{\pi}}\qty(\sqrt{\beta-i t}-\sqrt{\beta + it})
    \end{aligned}
    \\
    &\kappa^{3}_{\beta}(t)=\frac{-2 t^8 \sqrt{\beta ^2+t^2}}{45 \pi }-\frac{1658 \beta ^2 t^6 \sqrt{\beta^2+t^2}}{63 \pi }+\frac{16 i \beta ^2 t^7 \left(\pi -4 \arctan\left(\frac{\sqrt{\beta +i t}}{\sqrt{\beta -i t}}\right)\right)}{3\pi}\\
    &\begin{aligned}
        \kappa_\beta^{\frac 7 2}(t)=&\frac{i t^7 \sqrt{\beta ^2+t^2}}{210 \sqrt{\pi }} \left(\qty(756 \beta ^2-t^2) \left(\sqrt{\beta +i t}-\sqrt{\beta -i t}\right)+31 i \beta  t \left(\sqrt{\beta -i t}+\sqrt{\beta +i t}\right)\right)\\
        &+\frac{2048}{105} \sqrt{\frac{2}{\pi }} \beta ^{5/2} t^8.
    \end{aligned}
    \label{eq:k7/2}
\end{align}
From these results, one can observe that the contributions to the SFF at half-integer genus are mainly of the same functional form as the orientable contributions (cf. e.g. \cite{Saad2022,Blommaert2022,Weber2022}), i.e. exhibiting mostly a polynomial dependence on $\beta$ and $t$ with additional appearances in the form of $\sqrt{\beta}$, $\sqrt{\beta\pm i t}$ and $\sqrt{\beta^2+t^2}$. For integer genus however, there are $\arctan$-terms which do not occur in the GUE-like theory. Rewriting the $\arctan$ on $\mathbb{C}$ as a logarithm\footnote{A more detailed treatment of this type of terms is given in \cref{sec:arctan}.}, one finds
\begin{align}
    \arctan\qty(\sqrt{\frac{\beta + i t}{\beta - it}})=\frac{1}{2 i}\qty[\log\qty(\frac{-t+\sqrt{t^2+\beta^2}}{\beta})+i\frac\pi 2].
\end{align}
We first note that for all the cases we have computed, the $-i \frac \pi 2$ cancels all the other imaginary terms in the above contributions. This is necessary since by definition, the canonical SFF has to be a real function.

Since we are mainly interested in the behaviour of the genus expansion at large times, $\frac{\beta}{t}$ is a small quantity in the regime of interest. This motivates an expansion of the contributions in $\frac \beta t$ around $0$. Taking only the leading order behaviour, one finds
\begin{align}
    \frac{1}{2 i}\log\qty(\frac{-t+\sqrt{t^2+\beta^2}}{\beta})\rightarrow\frac{\log\qty(\frac{\beta}{2t})}{2 i}.
\end{align}
For \eqref{eq:k0}-\eqref{eq:k7/2}, this yields
\begin{align}\label{eq:g=0}
    \kappa^{0}_{\beta}(t)&\rightarrow \frac{t}{2\pi \beta}\\
    \kappa^{\frac 12}_{\beta}(t)&\rightarrow -\frac{t^2}{\sqrt{2\pi \beta }}\label{eq:g=1/2}\\
    \kappa^{1}_{\beta}(t)&\rightarrow -\qty[\frac{10}{3}+\log\qty(\frac{\beta}{2t})]\frac{t^3}{\pi}\label{eq:g=1}\\
    \kappa^{\frac 3 2}_{\beta}(t)&=-\frac{\sqrt{2\pi}t^{\frac 9 2}}{3\pi}+\frac{8\sqrt{2\pi \beta}}{3\pi} t^4 \label{eq:g=3/2}\\
    \kappa^{2}_{\beta}(t)&\rightarrow \frac{\beta t^5}{\pi}\qty[\frac{163}{15}+4 \log\qty(\frac{\beta}{2t})]\label{eq:g=2}\\
    \kappa^{\frac 5 2}_{\beta}(t)&\rightarrow-\frac{64\qty(2\pi\beta)^{\frac 3 2}}{15\pi^2} t^6+\frac{t^7\sqrt{t}}{15\sqrt{2\pi}}+\frac{17 \beta t^6 \sqrt{t}}{6\sqrt{2\pi}}\label{eq:g=5/2}\\
    \kappa^{3}_{\beta}(t)&\rightarrow\frac{-2 t^9}{45 \pi }-\frac{8297 \beta ^2 t^7}{315 \pi }-\frac{32 \beta^2 t^7 \log\qty(\frac{\beta}{2t})}{3\pi}
    \label{eq:g=3}\\
    \kappa^{\frac 7 2}_{\beta}(t)&\rightarrow \frac{t^{10}\sqrt{t}}{105 \sqrt{2 \pi }}-\frac{3 \beta  t^{9}\sqrt{t}}{10 \sqrt{2 \pi }}-\frac{881 \beta ^2 t^{8}\sqrt{t}}{120 \sqrt{2 \pi }}+\frac{2048}{105} \sqrt{\frac{2}{\pi }} \beta ^{5/2} t^8.\label{eq:g=7/2}
\end{align}
Comparing this expression to the corresponding expression of the GUE-like theory, we see the admission of unorientable surfaces leads to several differences. The first, and perhaps least interesting, is the inclusion of even powers of $t$ at half-integer genus. However, a rather striking difference is the appearance of terms having a logarithmic dependence on $\frac{2t}{\beta}$ for integer genus contributions. Terms of the same form have been discovered in a diagrammatic treatment of the orientable Airy model in \cite{Saad2022} as well, though they all cancelled. Tracing their origin to the unorientable Airy WP volumes one can see that they originate only from terms in $V^>_g$ that are of odd order in both lengths\footnote{For a derivation of this and a more detailed study of the structure of the contributions to $\kappa_\beta(t)$ the reader is referred to \cref{sec:SFF_general}.}. Recalling that the combined order of lengths in $V^>_g$ is given by $6g-2$, which for half-integer genus is odd while being even for integer genus, one can directly see that odd/odd terms, and thus the appearance of the logarithms, is only possible for integer genus.

A further difference compared to the SFF from the orientable theory can be seen after $\tau-$scaling, i.e. using $t=e^{S_0} \tau$. In the orientable theory, as first shown in \cite{Saad2022}, the $\tau-$scaled SFF can be written as 
\begin{align}
    \kappa^{\text{GUE}}_\beta(e^{S_0}\tau)=e^{S_0} f(\tau)
\end{align}
with $f$ having no dependence on $S_0$. Doing this for the unorientable SFF, e.g. the contribution at $g=\frac 3 2$, one finds 
\begin{align}
    \kappa_\beta(e^{S_0}\tau)
    & \supset e^{S_0}\sqrt{\beta}\tau^4 \qty(-\frac{\sqrt{2\pi}}{3\pi}\sqrt{\frac{t}{\beta}}+\frac{8\sqrt{2\pi}}{3\pi}).
\end{align}
This contains a term scaling as $\tau^4 \sqrt{t}$. Higher order square root terms occur for every half-integer genus we have computed starting at $g= \frac 3 2$.  From the $g=3$ contribution we see a term that, after $\tau-$scaling, scales as $\tau^7 t^2$. In the corresponding orientable theory no such terms appears, and the $\tau-$scaled SFF has no $t$ dependence (or terms of higher order than $1$ in $e^{S_0}$). In section \ref{sec:comparison} we argue that terms higher order in $t$, including the logarithms, are in fact related and can be eliminated using certain identities.

\section{Canonical SFF from universal RMT }\label{sec:RMTUniversality}
The aim of this section is to derive the canonical SFF from the perspective of universal RMT. The canonical SFF will then be computed for the orthogonal symmetry class and a leading order energy density corresponding to the Airy model.
\subsection{The universal RMT form of the canonical SFF}
The connected canonical two-point correlation function is given by a double Laplace transform of the connected density-density correlation function:
\begin{align}
\langle Z(\beta_1)Z(\beta_2)\rangle_c=\int_{0}^{\infty}\int_{0}^{\infty}\langle\rho^\text{T}(E_1)\rho^\text{T}(E_2)\rangle_c e^{-\beta_1 E_1-\beta_2 E_2}\dd{E_1}\dd{E_2},
\end{align}
where the superscript $\text{T}$ indicates ``total'', consistent with the notation of \cite{Saad2019}, e.g. $\langle \rho^T(E) \rangle $ = $e^{S_0}\langle \rho(E) \rangle $.
The SFF is then given by taking $\beta_1 = \beta + it$ and $\beta_2 = \beta -it$:

\begin{equation}
\begin{aligned}
\langle Z(\beta+it)Z(\beta-it)\rangle_c&=\int_{0}^{\infty}\int_{0}^{\infty}\langle\rho^{\text{T}}(E_1)\rho^{\text{T}}(E_2)\rangle_c e^{-\beta \qty(E_1+E_2)-it\qty(E_1-E_2)}\dd{E_1}\dd{E_2}\\
&=2\int_{0}^{\infty}\dd E e^{-2\beta E}\int_{0}^{2E}\dd{\Delta}\cos(t\Delta)\langle\rho^\text{T}(E+\frac\Delta 2)\rho^{\text{T}}(E-\frac\Delta 2)\rangle_c,
\end{aligned}
\end{equation}
where in the second line a change of variables to $E=\frac{E_1+E_2}{2}$, $\Delta=\abs{E_1-E_2}$ was performed.
To study the universal behaviour of the SFF, it needs to be evaluated in the late time limit, which we take to be of the order $e^{S_0}$. To accomplish this one can define a scaled time \cite{Saad2022}:
\begin{align}
	\tau = e^{-S_0}t,
 \label{eq:tau_sc}
\end{align}
where $\tau$ is assumed to be finite. Then one obtains
\begin{equation}
	\begin{aligned}
	&\langle Z(\beta+i\tau e^{S_0})Z(\beta-i\tau e^{S_0})\rangle_c\\
 &=2\int_{0}^{\infty}\dd E e^{-2\beta E}\int_{0}^{2E}\dd{\Delta}\cos(e^{S_0}\tau\Delta)\langle\rho^\text{T}(E+\frac\Delta 2)\rho^\text{T}(E-\frac\Delta 2)\rangle_c\\
	&=2e^{2S_0}\int_{0}^{\infty}\dd E e^{-2\beta E}\int_{0}^{2E}\dd{\Delta}\cos(e^{S_0}\tau\Delta)\langle\rho(E+\frac{\Delta}{2})\rho(E-\frac{\Delta}{2})\rangle_c, \\
 &=2e^{S_0}\int_{0}^{\infty}\dd E e^{-2\beta E}\int_{0}^{2Ee^{S_0}}\dd{x}\cos(\tau x)\langle\rho(E+\frac{x e^{-S_0}}{2})\rho(E-\frac{x e^{-S_0}}{2})\rangle_c.
	\end{aligned}
    \label{eq:kappa(tau)}
\end{equation}
In the second line the factors of $e^{S_0}$ in the total densities were factored out and in the last line the variable $x= e^{S_0} \Delta$ was defined.
The limit $e^{S_0} \to \infty$ is precisely what is known as the ``universal'' regime of the two-point function of the energy density which is known for all 10 Altland-Zirnbauer classes \cite{Gnutzmann2004}. To make the meaning of this limit precise, following the notation from \cite{Haake2010} adapted to the double scaled theory\footnote{i.e. $N\to e^{S_0}$.}, one defines the spectral correlation functions as:
\begin{align}
&R_1(E)=e^{S_0}\langle\rho(E)\rangle\\
&R_2(E_1,E_2)=e^{2S_0}\langle\rho(E_1)\rho(E_2)\rangle-e^{S_0}\delta(E_1-E_2)\langle\rho(E_1)\rangle,
\end{align}
and accordingly the connected two-point  correlation function\footnote{In \cite{Haake2010}, $\langle\rho(E_1)\rho(E_2)\rangle_c$ is referred to as $S(E_1,E_2).$}:
\begin{align}
C_2(E_1,E_2)&=R_1(E_1)R_1(E_2)-R_2(E_1,E_2)\\
&=e^{2S_0}\langle\rho(E_1)\rangle\langle\rho(E_2)\rangle+e^{S_0}\delta(E_1-E_2)\langle\rho(E_1)\rangle-e^{2S_0}\langle\rho(E_1)\rho(E_2)\rangle \\
&=e^{S_0}\delta(E_1-E_2)\langle\rho(E_1)\rangle-e^{2S_0}\langle\rho(E_1)\rho(E_2)\rangle_c.
\end{align}
The universal limit is then taken to be
\begin{align}
	\lim\limits_{e^{S_0}\to\infty}\frac{C_2(E_1,E_2)}{R_1(E_1)R_2(E_2)}=\Upsilon\qty(\expval{\rho(E)}x),\label{eq:def_upsilon}
\end{align}
such that $x$ is kept finite. The function $\Upsilon$ is referred to as the cluster function and depends on the specific symmetry class of the ensemble. The limit also justifies the replacement of $\langle\rho(E_1)\rangle=\langle\rho(E_2)\rangle=\langle\rho(E)\rangle$. Using the above explicit  expressions one can write,
\begin{align}
	\frac{C_2(E_1,E_2)}{R_1(E_1)R_2(E_2)}=\delta\qty(\ev{\rho(E)}x)-\frac{\langle\rho(E_1)\rho(E_2)\rangle_c}{\langle\rho(E)\rangle^2},
\end{align}
where one can take the universal limit and obtain the sought for expression:
\begin{align}
	\langle\rho(E+\frac{x e^{-S_0}}{2})\rho(E-\frac{x e^{-S_0}}{2})\rangle_c &= \delta\qty(x)\langle\rho(E)\rangle-\ev{\rho(E)}^2 \Upsilon\qty(\expval{\rho(E)}x).\label{eq:univers_rhorho}
\end{align}
We then define the universal limit of the canonical SFF in the following way:
\begin{align}
    e^{-S_0}\kappa_{\beta}(\tau) \coloneqq  \lim_{e^{S_0} \to \infty} e^{-S_0}\langle Z(\beta+i\tau e^{S_0})Z(\beta-i\tau e^{S_0})\rangle_c.
\end{align}
To take this limit we note that $\langle \rho(E) \rangle$ has a genus expansion in powers of $e^{-S_0}$. The leading order contribution, $\rho_0(E)$, is the genus zero term, such that
\begin{align}
    \rho_0(E) = \lim_{e^{S_0} \to \infty}\langle \rho(E) \rangle.
\end{align}
Therefore, the following replacement can be made in~\eqref{eq:univers_rhorho}:
\begin{align}
    \langle \rho(E) \rangle \to \rho_0(E).
\end{align}
Then~\eqref{eq:univers_rhorho} can be plugged into~\eqref{eq:kappa(tau)} and the upper limit of the integral over $x$ can be taken to infinity.\footnote{It is not at all obvious the corrections arising from keeping the upper bound finite are subleading in $e^{S_0}$, but we will not pursue this computation here.}
These considerations lead to the following definition of the universal canonical SFF:
\begin{align}
  e^{-S_0}\kappa_{\beta}(\tau) &=\int_{0}^{\infty}\dd E e^{-2\beta E}\rho_0(E)- 
2\int_{0}^{\infty}\dd E e^{-2\beta E}\rho_0(E)\int_{0}^{\infty}\dd{x} \cos(\frac{\tau}{\rho_0(E)}x)\Upsilon\qty(x)
\label{eq:SFF1}
\end{align}
where $\Upsilon(x)$, \cref{eq:def_upsilon}, depends on the specific symmetry class. 
\subsection{The Airy canonical SFF for the orthogonal symmetry class}
\label{sec:SFF10}
It is necessary to compute the universal canonical SFF, within the orthogonal symmetry class, for the Airy model using~\eqref{eq:SFF1} in order to compare it to the result derived using the unorientable Airy WP volumes. Since the Airy model is a low energy limit of JT gravity, this comparison provides a non-trivial test of the quantum chaotic nature of unorientable JT gravity. 
Specifically, the leading order contribution to the energy density for the Airy model is
\begin{equation}
    \rho^{\text{Airy}}_0(E) = \frac{\sqrt{E}}{2\pi}\label{eq:rho_0_def}.
\end{equation}
The cluster function for the orthogonal symmetry class then reads \cite{Mehta2004}:
\begin{align}
    \Upsilon(x)&=\frac{\sin[2](\pi x)}{\pi^2 x^2}+\qty(\int_x^\infty \dd t \frac{\sin(\pi t)}{\pi t})\qty(\dv{x}\frac{\sin(\pi x)}{\pi x})\\
    &=\frac{\sin[2](\pi x)}{\pi^2 x^2}+\qty(\frac 12 -\int_0^x \dd t \frac{\sin(\pi t)}{\pi t})\qty(\dv{x}\frac{\sin(\pi x)}{\pi x}).
\end{align}
The integral over $x$ in~\eqref{eq:SFF1} is the form factor \cite{Mehta2004}:
\begin{align}
    2\int_{0}^{\infty}\dd{x} \cos(\frac{\tau}{\rho_0(E)}x) \Upsilon\qty(x)=\left\{
	\begin{array}{ll}
		1-\frac{\tau}{\pi \rho_0(E)}+\frac{\tau}{2\pi \rho_0(E)}\log\qty(1+\frac{\tau}{\pi\rho_0(E)})  & \mbox{if } \frac{\tau}{2\pi} \leq \rho_0(E) \\
		-1+\frac{\tau}{2\pi\rho_0(E)}\log\qty(\frac{\frac{\tau}{\pi}+\rho_0(E)}{\frac{\tau}{\pi}-\rho_0(E)}) & \mbox{if } \frac{\tau}{2\pi} \geq \rho_0(E)
	\end{array}
\right.
\label{eq:cf}
\end{align}
The canonical SFF of the unorientable Airy model can then be computed by plugging~\eqref{eq:cf} into~\eqref{eq:SFF1} with $\rho_0(E)$ from \cref{eq:rho_0_def}:
\begin{equation}
    \begin{aligned}
        e^{-S_0}\kappa^{\text{GOE}}_{\beta}(\tau) =
        & \int_{0}^{\tau^2} \dd{E} e^{-2\beta E}\qty[\frac{\sqrt{E}}{\pi}-\frac{\tau}{2\pi}\log\qty(\frac{\tau}{\pi}+\frac{\sqrt{E}}{2\pi})+\frac{\tau}{2\pi}\log(\frac{\tau}{\pi}-\frac{\sqrt{E}}{2\pi})]+ \\[10pt]
        & \int_{\tau^2}^{\infty}\dd{E}e^{-2\beta E}\qty[\frac{\tau}{\pi}-\frac{\tau}{2\pi}\log\qty(1+\frac{2\tau}{\sqrt{E}})],
    \end{aligned}\label{eq:SFF_t_GOE_universal}
\end{equation}
where the superscript GOE indicates the universal RMT result for the orthogonal symmetry class. The solution, derived in Appendix \ref{rmt_derive}, reads:
\begin{equation}
\begin{aligned}\label{eq:SFF_GOE}
e^{-S_0}\kappa_{\beta}^{\text{GOE}}(\tau) =&{} \frac{1}{2\left(2\beta\right)^{3/2}\sqrt{\pi}}\Erf\left(\sqrt{2\beta \tau^2}\right)-\frac{\tau e^{-8\beta \tau^2}}{8\pi\beta}\Bigg{[}  \Gamma(0, 2\beta \tau^2)(1 - e^{8\beta \tau^2}) + \\[5pt]
& 16 \beta  \tau ^2 \, _2F_2\left(1,1;\frac{3}{2},2;8 \beta  \tau ^2\right)+\pi \Erfi\left(\sqrt{8\beta\tau^2 }  \right)-\\[5pt]& \sum_{n=1}^{\infty}\left(\sum_{m=1}^{n}\frac{(-1)^{n+m}(2)^{2m}}{(m)!(n-m)!(n-\frac{m}{2})}\right)\left(2\beta\tau^2\right)^n \Bigg{]}.
\end{aligned}
\end{equation}
Here ${}_2F_2$ is the generalized hypergeometric function and Erfi is the imaginary error function.
The first few terms of the expression are:
\begin{align}
&{} e^{-S_0}\kappa_{\beta}^{\text{GOE}}(\tau) = \nonumber \\[5pt]
&\frac{\tau}{2\pi \beta}-\frac{\tau^2}{\sqrt{2\pi\beta}}-\frac{\gamma +\log\qty(2\beta \tau^2)+\frac{1}{3}}{\pi}\tau^3+\frac{8\sqrt{2\pi \beta}}{3\pi}\tau^4+\frac{\beta\qty(4\gamma+4\log\qty(2\beta \tau^2)-\frac{7}{15} )}{\pi }\tau^5-\nonumber\\ 
&\frac{64\qty(2\pi\beta)^{\frac 32}}{15 \pi^2}\tau^6+\order{\tau^7},
\label{eq:SFF_rmt}
\end{align}
with $\gamma$ denoting the Euler-Mascheroni constant.
It remains to compare this expression to the result derived from the unorientable Airy WP volumes.

\section{Comparison of unorientable topological gravity and universal RMT}
\label{sec:comparison}
In this section we will explicitly demonstrate the non-trivial equivalence of the $\tau^3$ term in both expressions of the canonical SFF. It will also be shown that terms containing even powers of $\tau$ are in agreement for all terms so far computed, i.e. up to $g=\frac 7 2$\footnote{The terms with even powers of $\tau$ without $t-$dependence all agree. It remains to show the terms with $t-$dependence vanish.}. The reason for limiting the comparison of odd powers to $\tau^3$ lies in the fact that demonstrating this equivalence requires higher order terms in the loop equations than one would naively expect, e.g. the $g=4$ contribution for $\tau^5$. 

To compare the universal RMT result,~\eqref{eq:SFF_rmt}, to the result from the unorientable Airy WP volumes we first rewrite the latter as
\begin{align}
\label{eq:SFF3}
    e^{-S_0}\kappa^{\text{WP}}_{\beta}(t) = \sum_{g=0,\frac{1}{2},1  \dots} e^{-(2g+1)S_0}\kappa_{\beta}^{g}(t),
\end{align}
where $\kappa^{g}_{\beta}$ are given by \cref{eq:g=0,eq:g=1/2,eq:g=1,eq:g=3/2,eq:g=2,eq:g=5/2,eq:g=3,eq:g=7/2} and the superscript WP indicates this was computed from the unorientable Airy WP volumes. The aim is to show $\kappa^{\text{WP}}_{\beta}(\tau) = \kappa^{\text{GOE}}_{\beta}(\tau)$ in the universal limit, at least for the first few orders. The terms up to $g=\frac 7 2$ of~\eqref{eq:SFF3} after $\tau$ scaling, i.e. using $\tau=t e^{-S_0}$, are
\begin{align}
\begin{aligned}
 & e^{-S_0}\kappa^{\text{WP}}_{\beta}(\tau) = \frac{\tau}{2\pi \beta} -\frac{\tau^2}{\sqrt{2\pi \beta }}+ \frac{\tau^3}{\pi}\qty[\frac{-10}{3}+\log\qty(\frac{2t}{\beta})]-\frac{\sqrt{2\pi\beta}\tau^4}{3\pi}\left(\frac{t}{\beta}\right)^{1/2}\\[5pt]
  &+\frac{8\sqrt{2\pi \beta}}{3\pi} \tau^4 +\frac{\beta \tau^5}{\pi}\qty[\frac{163}{15}-4 \log\qty(\frac{2t}{\beta})]-\frac{64\qty(2\pi\beta)^{\frac 3 2}}{15\pi^2} \tau^6+\frac{17 \tau^6 \beta\sqrt{2\pi\beta}} {12\pi}\left(\frac{t}{\beta}\right)^{1/2}  \\[5pt]
  &+\frac{\tau^6\left(2\pi\beta\right)^{3/2}}{60 \pi^2}\left(\frac{t}{\beta}\right)^{3/2}+\frac{\beta^2\tau^7}{\pi}\qty[-\frac{8297}{315}+ \frac{32 }{3}\log\qty(\frac{2t}{\beta})]-\frac{2 \tau^7\beta^2}{45 \pi }\left(\frac{t}{\beta}\right)^2 \\
  &+\frac{\tau^8 \qty(2\pi \beta)^\frac{5}{2}}{840}\qty(\frac{t}{\beta})^{\frac{5}{2}}-\frac{3 \tau^8 \beta \qty(2\pi \beta)^\frac{3}{2}}{40 \pi^2}\qty(\frac{t}{\beta})^\frac{3}{2}-\frac{881 \tau^8\beta^2 \sqrt{2\pi\beta} }{240}\sqrt{\frac{t}{\beta}}+\frac{512(2\pi\beta)^{\frac 5 2}}{105\pi^3}  \tau^8 \\
  & + \order{\tau^9}\label{eq:kappa_WP}
  \end{aligned}
\end{align}
Note that after $\tau$ scaling there remain powers of $t$, however, the terms containing powers of $t$ can be shown to be a part of an asymptotic series, and in this way eliminated. In order to accomplish this elimination the series in \cref{eq:kappa_WP} has to be reorganized by ordering together terms of equal dependence on $\beta$, as suggested by the universal RMT result \cref{eq:SFF_rmt}: 
\begin{equation}
    \begin{aligned}
        & e^{-S_0}\kappa^{\text{WP}}_{\beta}(\tau) = \frac{\tau}{2\pi \beta} -\frac{\tau^2}{\sqrt{2\pi \beta }}+\frac{8\sqrt{2\pi \beta}}{3\pi} \tau^4-\frac{64\qty(2\pi\beta)^{\frac 3 2}}{15\pi^2} \tau^6 +\frac{512\qty(2\pi\beta)^{5/2}}{105\pi^3} \tau^8+\dots \\[7pt]
        &+\frac{\tau^3}{\pi} \qty[\frac{-10}{3}+\log\qty(\frac{2t}{\beta})-\frac{\sqrt{2\pi}}{3}\left(t\tau^2\right)^{1/2} +\frac{\sqrt{2\pi}}{30}\left(t \tau^2\right)^{3/2}-\frac{2\left(t\tau^2\right)^2}{45}+\frac{\sqrt{2\pi}}{210}(\tau^2 t)^{5/2}+\dots]\\[7pt] 
        &+\frac{\tau^5\beta}{\pi}\qty[\frac{163}{15}-4 \log\qty(\frac{2t}{\beta})+\frac{17 \sqrt{2\pi}} {12}\left(t\tau^2\right)^{1/2}-\frac{3\sqrt{2\pi}}{20}\left(t\tau^2\right)^{3/2} +\dots] \\[7pt]
        &+\frac{\tau^7 \beta^2}{\pi}\qty[-\frac{8297}{315}+\frac{32 }{3}\log(\frac{2 t}{\beta})-\frac{881\sqrt{2\pi}}{240}(\tau^2 t)^{1/2}+\dots] \\[7pt]
        &+\dots.
 \label{eq:SFF4} 
    \end{aligned}
\end{equation}
Here, we have added dots to make precise where additional terms would appear upon computing volumes of higher genus. Specifically, in the first line the dots indicate terms of even power in $\tau$ and half integer power in $\beta$ that appear as the $t$-independent terms from half-integer genus contributions. In the subsequent lines, we group together terms having the same dependence on integer powers of $\beta$. The dots in the groupings of these terms mean $t$-dependent terms of the same $\beta$-dependence from higher genus. Here, the type of additional terms is unambiguously two-fold. First, there are terms depending on $\sqrt{t}$, which come from half-integer genus contributions.
Secondly, there are terms that do not depend on $\sqrt{t}$ but on $t$, for which there is only one example available at present, given by the fifth term in the second line of \cref{eq:SFF4}, arising from the $g=3$ contribution. We believe both these types of terms will continue for higher genus, thus the dots. 
The dots in the last line of \cref{eq:SFF4} indicate terms of higher order in $\beta$ and $\tau$ that definitely arise at higher genus, again with the sort of $t$-dependent terms discussed. 

Comparing~\eqref{eq:SFF4} and~\eqref{eq:SFF_rmt}, one can see that all of the terms with even powers of $\tau$ agree if we exclude the terms with $t-$dependence.  The coefficients of $\log(\beta)$ also agree for all terms computed. However, it is not at all obvious that the coefficients of odd powers of $\tau$ agree due to them being dependent on $t$ which is not the case for the universal RMT result. Furthermore, it is not obvious how the $t-$dependent terms with even powers of $\tau$, which we have included in the brackets, vanish. We will present a program to systematically remove the $t$-dependent terms from~\eqref{eq:SFF4}, and in doing so demonstrate the equivalence of the $\tau^3$ term to the result from universal RMT. 

To prove the $\tau^3$ coefficient of $\kappa^{\text{WP}}_{\beta}(\tau)$, as written in~\eqref{eq:SFF4}, agrees with the  $\tau^3$ coefficient of $\kappa^{\text{GOE}}_{\beta}(\tau)$, given by~\eqref{eq:SFF_rmt}, consider the following generalized hypergeometric functions: 
\begin{align}
\begin{aligned}
&\frac{\tau^3\left(t\tau^2\right)^2}{45\pi}\left(6{}_2F_2\left(2,2;3,\frac{7}{2};-t\tau^2\right) - 4{}_1F_1\left(\frac{3}{2};\frac{7}{2};\frac{-t\tau^2}{2}\right)\right) \\[10pt]
&= \frac{2\tau^7 t^2}{45\pi}+\underbrace{\frac{\tau^3\left(t\tau^2\right)^2}{45\pi} \sum_{k=1}^{\infty}a_k (t \tau^2)^k}_{\order{\tau^9}}.
   \label{eq:SFF5}
\end{aligned}
\end{align}
The coefficients $a_k$ can be read off from the definition of hypergeometric functions but will not be important for our purposes. The important point is that the terms in the series are $\order{\tau^9}$, i.e. at least $g=4$ contributions, and are independent of $\beta$.
Then consider the asymptotic expansion of these functions \cite{NIST:DLMF}:
\begin{align}
\begin{aligned}    &\frac{\tau^3\left(t\tau^2\right)^2}{45\pi}\left(6{}_2F_2\left(2,2;3,\frac{7}{2};-t\tau^2\right) - 4{}_1F_1\left(\frac{3}{2};\frac{7}{2};\frac{-t\tau^2}{2}\right)\right)  \\[10pt]
& \overset{t \to \infty}{=} \frac{\tau^3}{\pi}\left(-\frac{\sqrt{2\pi}}{3}\left(t\tau^2\right)^{1/2} +\log \left(4 t \tau ^2\right)+\gamma-3 \right) .
    \label{eq:SFF6}
\end{aligned}
\end{align}
We note that~\eqref{eq:SFF6} has terms such as $\order{t^{-1/2}}$ that we are dropping in the limit $t \to \infty$. Setting the convergent series~\eqref{eq:SFF5} equal to the asymptotic series~\eqref{eq:SFF6}, and adding zero in the form of $\log(\frac{\beta}{\beta})$, we see that: 

\begin{align}
    - \tau^3\frac{\log \left(2\beta  \tau ^2\right)+\gamma+\frac{1}{3}}{\pi} \overset{t\to\infty}{=}  &  \frac{\tau^3}{\pi}\left(-\frac{10}{3}+\log \left(\frac{2t}{\beta}\right) -\frac{\sqrt{2\pi}}{3}\left(t\tau^2\right)^{1/2} -\frac{2\left(t\tau^2\right)^2}{45}\right)-\order{\tau^9} .
    \label{tau^3}
\end{align}
Note the coefficient of $\tau^3$ on the left hand side is the same as the result in $\kappa^{\text{GOE}}_{\beta}(\tau)$ from~\eqref{eq:SFF_rmt}. We repeat the bracketed term associated with $\tau^3$ in~\eqref{eq:SFF4}:
\begin{align}
    \frac{\tau^3}{\pi} \qty[\frac{-10}{3}+\log\qty(\frac{2t}{\beta})-\frac{\sqrt{2\pi}}{3}\left(t\tau^2\right)^{1/2} +\frac{\sqrt{2\pi}}{30}\left(t \tau^2\right)^{3/2}-\frac{2\left(t\tau^2\right)^2}{45}+\frac{\sqrt{2\pi}}{210}(\tau^2 t)^{5/2}+\dots]
 \label{brackets1}
\end{align}
Using the identity~\eqref{tau^3}, we can recover the $\tau^3$ result in the limit $t \to \infty$ as given by universal RMT~\eqref{eq:SFF_rmt}:
\begin{align}
    - \tau^3\frac{\log \left(2\beta  \tau ^2\right)+\gamma+\frac{1}{3}}{\pi}+ \underbrace{\qty[ \frac{\tau^3}{\pi}\frac{\sqrt{2\pi}}{30}\left(t \tau^2\right)^{3/2}+\frac{\tau^3}{\pi}\frac{\sqrt{2\pi}}{210}(\tau^2 t)^{5/2}-\order{\tau^9}+\dots]}_{\order{\tau^6}}.
    \label{brackets2}
\end{align}
This demonstrates the equivalence at the order $\tau^3$ because the term in brackets is $\order{\tau^6}$. Furthermore, it proves the equivalence at the order $\tau^4$ because we have eliminated the $\sqrt{t}-$dependent $\tau^4$ term in~\eqref{brackets1}, and the first line in $\eqref{eq:SFF4}$ gives the correct $\tau^4$ term. 

Now since the terms in brackets in~\eqref{brackets2} are independent of $\beta$ all of these terms must vanish. To prove the equivalence of terms $\tau^6$ and higher it will be necessary to show this, i.e. to prove:
\begin{align}
    \lim_{t \to \infty}\qty[ \frac{\tau^3}{\pi}\frac{\sqrt{2\pi}}{30}\left(t \tau^2\right)^{3/2}+\frac{\tau^3}{\pi}\frac{\sqrt{2\pi}}{210}(\tau^2 t)^{5/2}-\frac{\tau^3\left(t\tau^2\right)^2}{45\pi} \sum_{k=1}^{\infty}a_k(t\tau^2)^k+\dots]= 0
    \label{brackets3}
\end{align}
where we have reinserted the series from~\eqref{eq:SFF5} and we note again that the $a_k$ are known constants.
We argue that the program we have introduced, taking advantage of asymptotic expansions of hypergeometric functions, can be used to show all these terms vanish order by order in $\tau$. First note that the dots in~\eqref{brackets3} begin at $g=4$, i.e. $\order{\tau^9}$, which we have not yet computed. We expect these terms to continue with the same structure, that is a series of terms proportional to even powers of $\tau$ and $\sqrt{t}$ and terms proportional to odd powers of $\tau$ and integer powers of $t$. We have already seen from~\eqref{eq:SFF6} and~\eqref{eq:SFF5} how the term $\propto \tau^4 \sqrt{t}$ came from the asymptotic expansion of hypergeometric functions and the term $\propto \tau^7 t^2$ came from the convergent expansion. Consequently, it is plausible that one can derive an analogous identity to~\eqref{tau^3} to cancel the terms in~\eqref{brackets3} $\propto \tau^6 t^{3/2}$ and $\propto \tau^8 t^{5/2}$ against the term $\propto \tau^9 t^3$. The necessary functions to do this would have to have an asymptotic expansion starting at $\propto \tau^6 t^{3/2}$ and a convergent expansion beginning at $\propto \tau^9 t^3$. To derive the exact identity, we need the $g=4$ term from the volumes to know for certain what the term $\propto \tau^9 t^3$ exactly is. We only know that part of this term will be from the $k=1$ term in~\eqref{brackets3}, but it is possible the ``dots'' also contribute to it. However, we can show the necessary functions do exist. For example, an appropriate combination of,
\begin{align}
    \tau^3(t\tau^2)^3{}_1F_1\left(3/2,5/2;-\frac{t\tau^2}{2}\right) \hspace{.3cm}\text{and}\hspace{.3cm}  \tau^3(t\tau^2)^3{}_1F_1\left(1/2,5/2;-\frac{t\tau^2}{2}\right)
\end{align}
should work, though it is possible a slight variation of the functions' arguments are needed. Of course, this would only eliminate terms up to order $\tau^9$, and more identities are required for each higher order in $\tau$, but the procedure can be continued indefinitely.

So far we have shown the equivalence of $\kappa^{\text{WP}}_{\beta}\left(\tau\right)$ and $\kappa^{\text{GOE}}_{\beta}\left(\tau\right)$ up to $\tau^4$, i.e. $g=3/2$, and have argued that the remaining $\beta$ independent terms must vanish. To extend the agreement to $\tau^5$ we have to analyze the bracket proportional to $\beta$ in~\eqref{eq:SFF4}, which is
\begin{align}
    &\frac{\tau^5\beta}{\pi}\qty[\frac{163}{15}-4 \log\qty(\frac{2t}{\beta})+\frac{17 \sqrt{2\pi}} {12}\left(t\tau^2\right)^{1/2}-\frac{3\sqrt{2\pi}}{20}\left(t\tau^2\right)^{3/2} +\dots]
    \label{brackets6}
\end{align}
We again note that the coefficient of $\tau^5\log(\beta)$ in~\eqref{brackets6} agrees with the universal RMT result~\eqref{eq:SFF_rmt}. For this reason and because the structure of the terms is the same as~\eqref{brackets1}, the same exact argument made for $\tau^3$ and the $\beta$ independent terms can be applied to the terms in~\eqref{brackets6}. Therefore, it should be possible to derive an analogous identity to~\eqref{tau^3} to demonstrate the equivalence of the $\tau^5$ term. However, without the $g=4$ contribution the exact identity cannot be derived, though the necessary functions certainly exist. It is seen that the structure of the terms in brackets continues for the terms proportional to $\beta^2$, so the same argument is applicable for the $\tau^7$ term as well.

To summarize, we believe the brackets associated with odd powers of $\tau$ and integer powers of $\beta$ in~\eqref{eq:SFF4} form an infinite series that, in the limit $t \to \infty$, goes to the universal RMT result~\eqref{eq:SFF_rmt}. The identity~\eqref{tau^3} gives a mechanism for how this occurs and demonstrates the equivalence for the $\tau^3$ and $\tau^4$ terms. To extend the program, i.e. cancel more $t$-dependent terms and extend agreement to higher order powers of $\tau$, we would need contributions from higher genus.

\section{Outlook: Cancellations}\label{sec:Cancellations}
Having presented evidence that the $\tau$ scaled limit of unorientable topological gravity agrees with the prediction of universal RMT for the orthogonal case, we can proceed to use this, in the spirit of \cite{Weber2022,Blommaert2022}, to infer constraints on the coefficients of the unorientable Airy WP volumes. To briefly recapitulate the idea and because we will need the result in a specific form for later comparison we briefly revisit the consideration from the unitary symmetry class of \cite{Weber2022}.

The first step one has to take is to compute the canonical SFF from the general structure of the respective Airy WP volumes, using general coefficients. In the unitary case the structure of the Airy WP volumes is given by \cref{eq:V_orientable}. Using the standard integration of trumpet partition functions $Z_t(\beta,b)$ against the volumes to obtain the canonical SFF one finds
\begin{align}
\begin{aligned}
    \kappa^{g,\upbeta=2}_\beta(t)&=\sum_{n+m=3g-1}C^{\upbeta=2}_{n,m}\frac{n!m!2^{2(n+m)}}{\pi}\qty(\beta+it)^{n+\frac 12}\qty(\beta -it)^{m+\frac 12}\\
    &=t^{3g}\sum_{n+m=3g-1} f(g) C^{\upbeta=2}_{n,m} n! m! (-1)^m\sum_{j,k=0}^\infty \binom{n+\frac 12}{j}\binom{m+\frac 12}{k}i^{k+j}\qty(-1)^j\qty(\frac{\beta}{t})^{k+j},
\end{aligned}
\end{align}
wherein the second line we used the generalized binomial theorem, assuming $t>\beta$ and introduced $f(g)$ to collect all the factors that depend only on the genus or are constants irrelevant to the following discussion. Summing over the contributions to the genus expansion including the weighting by $e^{2 S_0}$, performing the $\tau$-scaling, and ordering the summands by powers in $\tau$ one finds
\begin{align}
\begin{aligned}
    \kappa^{\upbeta=2}_{\beta}(\tau)=&\sum_{g=0}^{\infty}e^{g S_0}\tau^{3g}\sum_{n+m=3g-1}\sum_{l=0}^\infty C(g,l) e^{-l S_0}\tau^{-l}\beta^l \\
    & C^{\upbeta=2}_{n,m}n! m! (-1)^m \sum_{k+j=l}\binom{n+\frac 12}{k}\binom{m+\frac 12}{j}(-1)^j,\label{eq:SFF_GUE_expanded}
\end{aligned}
\end{align}
where $C(g,l)$ includes $f(g)$ in the equation above and also captures the dependence on $l$.
After having written the canonical SFF for the unitary symmetry class in this form, we come to the vital second part of our argument. Namely, as we can see from \cref{eq:SFF1} the RMT prediction, shown to agree with the topological gravity prediction in \cite{Saad2022}, has the form 
\begin{align}
    \kappa^{\upbeta=2}_{\beta}(\tau)=e^{S_0}g(\tau),
\end{align}
with $g(\tau)$ containing no dependence on $e^{S_0}$. This consequently implies that all terms in \cref{eq:SFF_GUE_expanded} being of higher order than $1$ in $e^{S_0}$ have to cancel. Specifically, defining
\begin{align}
    K^{\upbeta=2}_g(l)\coloneqq\sum_{n+m=3g-1} C^{\upbeta=2}_{n,m}n! m! (-1)^m \sum_{k+j=l}\binom{n+\frac 12}{k}\binom{m+\frac 12}{j}(-1)^j,
\end{align}
this amounts to the statement that for a given genus $g$
\begin{align}
    \underset{0\leq l<(g-1)}{\forall} K^{\upbeta=2}_g(l)=0.\label{eq:constraints_GUE}
\end{align}
These constraints can be cast into a much more convenient form as shown in \cite{Weber2022}. However, for the present purpose it is more useful to rewrite the constraints in a way that is better comparable to the orthogonal result. In order to motivate this, we recall that the unitary WP volumes only contain even powers of the lengths while all powers appear for the orthogonal case. Thus, to facilitate comparison to the orthogonal case it makes sense to write the constraints as
\begin{align}
    K^{\upbeta=2}_g(l)=\sum_{\substack{\alpha+\gamma=6g-2\\
    \alpha,\gamma \ \text{even}}} C^{\upbeta=2}_{\frac \alpha 2,\frac \gamma 2 }\Gamma\qty(\frac \alpha 2 +1)\Gamma\qty(\frac \gamma 2 +1) (-1)^{\frac \gamma 2}\sum_{k+j=l}\binom{\frac{\alpha+1}{2}}{k}\binom{\frac{\gamma+1}{2}}{j}(-1)^j.\label{eq:Constraint_GUE_oForm}
\end{align}
Having recalled the constraints on the coefficients in the unitary case, we now apply the concept to the orthogonal case.
There the structure of the Airy WP volumes is generalized to the form reported in \cref{eq:AiryWP_form_conjecture}. This generalization, compared to the unitary case, leads to two novel features in the large $t$ limit of the canonical SFF: The emergence of $\log\qty(\frac{2t}{\beta})$ at integer genus and powers of $t$ that survive the $\tau-$scaling limit. 
As we argued in the preceding chapter, by using identities derived from hypergeometric functions, these types of terms can be combined to produce the logarithm expected from universal RMT. Once these identities have been established, the cancellations can be studied in the same way as in the unitary case.
In the following, we present the first step of this study, considering only a subset of terms related to the logarithmic terms, while referring a complete study to future work.

The terms in the canonical SFF containing a $\log\qty(\frac{2t}{\beta})$ are particularly interesting because they do not appear in the unitary case. Thus the cancellations inferred for these terms are genuinely new. As we discussed above, logarithmic terms can only arise from monomials in the $V^>_g(b_1,b_2)$ that are odd in both lengths. As we show in \cref{sec:SFF_general}, the contribution of the term
\begin{align}
    b_1^{\alpha}b_2^{\gamma}\theta\qty(b_1-b_2)
\end{align}
to the canonical SFF is given by 
\begin{align}
    I(\alpha,\gamma)&=\frac{1}{8\pi\sqrt{\beta_1\beta_2}}\qty(-\pdv{}{a})^{\frac {\gamma+1} 2}\qty(-\pdv{}{b})^{\frac {\alpha+1} 2}\frac{\arctan\qty(\sqrt{\frac a b})}{\sqrt{ab}},
\end{align}
with $a=\frac{1}{4\beta_2}$, $b=\frac{1}{4\beta_1}$. As we are only interested in the logarithmic terms which originate from the $\arctan$ we extract the part from $I(\alpha,\gamma)$ that contains an $\arctan$, leading to
\begin{align}
\begin{aligned}
     I(\alpha,\gamma)&\rightarrow \frac{2^{6g-1}\Gamma\qty(1+\frac \alpha 2)\Gamma\qty(1+\frac\gamma 2)}{\pi^2}\beta_1^{\frac{\alpha+1}{2}}\beta_2^{\frac{\gamma+1}{2}}\arctan\qty(\sqrt{\frac{\beta_1}{\beta_2}})\\
     &\eqqcolon I^{\log}_{\alpha,\gamma}(\beta_1,\beta_2).
\end{aligned}
\end{align}
Plugging into this expression, as for the unitary case $\beta_1=\beta+i t=\beta_2^*$ and summing over all the contributing monomials in $V_g^>$, including also the complementary $\theta$-function, one finds
\begin{align}
    \begin{aligned}
        \kappa^{\upbeta=1}_\beta(t)\supset \kappa^{\log}_\beta(t)\coloneqq\sum_{g=0}^\infty e^{-2g S_0}\sum_{\substack{\alpha+\gamma=6g-2\\ \alpha,\gamma  \text{ odd}}}I^{\log}_{\alpha,\gamma}(\beta_1,\beta_2)+ \qty(\beta_1\leftrightarrow\beta_2).
    \end{aligned}
\end{align}
One can write the contribution at genus $g$ to the first part of the sum as 
\begin{align}
\begin{aligned}
    e^{-2g}&\arctan\qty(\sqrt{\frac{\beta+ it}{\beta -it }})\frac{t^{3g}}{\pi}\sum_{l=0}^{3g}C(g,l)\beta^l t^{-l}\\
    &\underbrace{\sum_{\substack{\alpha+\gamma=6g-2\\ \alpha,\gamma  \text{ odd}}}C_{\alpha,\gamma}\frac{\Gamma(1+\frac \alpha 2)\Gamma(1+\frac \gamma 2)}{\pi} (-1)^{\frac{\gamma+1}{2}}
    \sum_{n+m=l}\binom{\frac{\alpha+1}{2}}{n}\binom{\frac{\gamma+1}{2}}{m}(-1)^m}_{\coloneqq K_g^{\upbeta=1}(l)}.
\end{aligned}
\end{align}
Here we defined $C(g,l)$ to be a non-vanishing function depending only on $g$ and $l$ and thus being irrelevant for the present discussion. Furthermore, we note that $K_g^{\upbeta=1}(l)$ is a rational number because the $C_{\alpha,\gamma}\in \mathbb{Q}_+$ by definition, the binomial coefficients are also natural numbers as $\alpha$ and $\gamma$ are odd, and the factor of $\pi$ in the denominator is cancelled by the two factors of $\sqrt{\pi}$ arising from the $\Gamma$-functions due to both indices being odd.
Expanding now the $\arctan$ for large times, as we prove in \cref{sec:arctan}, one finds that for even powers in $t$ multiplying it, effectively
\begin{align}
    t^{\text{even}} \arctan\qty(\sqrt{\frac{\beta+ it}{\beta -it }}) \rightarrow \frac \pi 2,
\end{align}
and for odd powers 
\begin{align}
   t^{\text{odd}} \arctan\qty(\sqrt{\frac{\beta+ it}{\beta -it }}) \rightarrow \log\qty(\frac{\beta}{2t})+\sum_{k=0}^{\infty}C(2k)\qty(\frac{\beta}{t})^{2k}.
\end{align}
Here the $C(k)$ are rational constants defined in \cref{sec:arctan} and ``$\rightarrow$'' denoting that the rhs is the contribution to the canonical SFF arising from the contribution from the lhs upon including also the complementary contribution to it, i.e. the one stemming from the $\theta(b_2-b_1)$-part of the unorientable Airy WP volume. Thus, if after $\tau$-scaling the $t$ multiplying the $\arctan$ yield a contribution of higher order in $e^{S_0}$ than 1, the contribution to the $\tau$-scaled canonical SFF from this term is 
\begin{align}
    e^{(g-l)S_0}K_g^{\upbeta=1}(l)\frac{\tau^{3g-l}}{\pi}
            \begin{cases}
                \log\qty(\frac{\beta}{2t})+\sum_{k=0}^{\infty}C(2k)\qty(\frac{\beta}{t})^{2k} &\text{for}\ 3g-l \ \text{odd}, \\
                \frac{\pi}{2}\quad &\text{for} \ 3g-l \ \text{even}. 
          \end{cases}
\end{align}
Having computed the general contribution to the $\tau$-scaled canonical SFF, the next step is to compare to the prediction from universal RMT.
At first glance this would amount to requiring all contributions at higher order than $1$ in $e^{S_0}$ to cancel, as for the unitary case. Notably, this is not as easily achieved as in the latter case, since by the considerations given in \cref{sec:comparison} also terms of higher order than $1$ in $e^{S_0}$ in the result from the unorientable Airy WP volumes can be transformed to agree with the universal RMT result. However, the transformation does not cancel terms containing $\log\qty(\frac{\beta}{2t})$ as those are mapped to a logarithm of different argument and would persist. This implies, that for the case of odd $g-l>1$, $K_g^{\upbeta=1}(l)$ has to vanish. Furthermore, the structure in terms of factors of $\pi$, namely that at integer genus the contribution to the canonical SFF is of order $\pi^{-1}$, can only be transformed into terms of order $\pi^{-\frac 12}$. Thus, terms that are of order $\pi^0$ would transform to terms of order $\pi^{\frac 12}$, incompatible with the universal RMT result. Curiously, for the present case, agreement with the universal RMT result does require all the terms comprising a higher order than $e^{S_0}$ to vanish, i.e.
\begin{align}
    \underset{0\leq l<(g-1)}{\forall} K^{\upbeta=1}_g(l)=0.\label{eq:constraints_GOE}
\end{align}
Having claimed this, we will give some examples. As the type of terms under consideration appear only for integer genus we will consider all the cases available for us so far, i.e. $g=1,2,3$. For genus one there is of course no cancellation as the maximal order in $e^{S_0}$ is given by $g$ which here of course is one already. For the case of $g=2$ one finds one constraint which one can work out to be
\begin{align}
   K^{\upbeta=1}_2(0)\propto 21 C_{1,9}-7 C_{3,7}+5 C_{5,5}-7 C_{7,3}+21 C_{9,1}.
\end{align}
For $g=3$ there are two, which are given by
\begin{align}
    K^{\upbeta=1}_3(0)&\propto 715 C_{1,15}-143 C_{3,13}+55 C_{5,11}-35 C_{7,9}+35 C_{9,7}-55 C_{11,5}+143 C_{13,3}-715 C_{15,1},\\
    K^{\upbeta=1}_3(1)&\propto 1001 C_{1,15}-143 C_{3,13}+33 C_{5,11}-7 C_{7,9}-7 C_{9,7}+33 C_{11,5}-143 C_{13,3}+1001 C_{15,1}.
\end{align}
Using the results for the respective volumes given in \cref{app:Volumes} one can read off
\begin{align}
    \begin{aligned}
        V_2^>(b_1,b_2)=\frac{1}{232243200}( &3885 b_1^{10}+58275 b_2^2 b_1^8+30720 b_2^3 b_1^7+186690 b_2^4 b_1^6+64512 b_2^5 b_1^5+\\
        &+183330 b_2^6 b_1^4+46080 b_2^7 b_1^3+38835 b_2^8 b_1^2+10240 b_2^9 b_1+1885 b_2^{10} ),
    \end{aligned}
\end{align}
\begin{align}
\begin{aligned}
    V_3^>(b_1,b_2)=&\frac{1}{2142493684531200}(887887 b_1^{16}+35515480 b_2^2 b_1^{14}+18350080 b_2^3 b_1^{13}+\\
    &+348079732 b_2^4 b_1^{12}+143130624 b_2^5 b_1^{11}+1242409168 b_2^6 b_1^{10}+449839104 b_2^7 b_1^9+\\
    &+1735393660 b_2^8 b_1^8+599785472 b_2^9 b_1^7+1049704656 b_2^{10} b_1^6+286261248 b_2^{11} b_1^5+\\
    &+269432436 b_2^{12} b_1^4+55050240 b_2^{13} b_1^3+21347416 b_2^{14} b_1^2+3670016 b_2^{15} b_1+447567 b_2^{16}).
\end{aligned}
\end{align}
Plugging he coefficients into the constraints one finds, as it should be, that they are indeed fulfilled.

Having shown these examples, we briefly compare $K^{\upbeta=1}_g(l)$ to their unitary counterparts which, specifically when written as in \cref{eq:Constraint_GUE_oForm}, are just the same, only that the sum runs over even $(\alpha,\gamma)$ as opposed to the sum over odd pairs in the orthogonal case. However, it is important to point out that while some of the unitary constraints are fulfilled directly due to the symmetry of the coefficients this is not the case for the orthogonal constraints which are all non-trivial. Nevertheless, the similarity is interesting and might point at a similar origin as the unitary constraints which have been shown to reflect properties of intersection numbers on the canonical SFF \cite{Blommaert2022}.

\section{Conclusion}
We have presented strong evidence of the quantum chaotic nature of unorientable topological gravity, as determined by the agreement of the $\tau$ scaled limit of the canonical SFF with universal RMT for the orthogonal symmetry class, i.e. corresponding to the universality class of systems with time-reversal invariance. Although such agreement has previously been established for JT gravity on orientable surfaces \cite{Saad2022}, generalizing this result to unorientable JT gravity faces extra difficulties due to the infinite volume of the unorientable WP volumes. For this reason, the agreement between RMT and the Airy limit considered here has important implications for unorientable JT gravity, as it makes it plausible that the $\tau-$scaled limit of the SFF for unorientable JT gravity agrees with universal RMT. However, the fidelity to RMT in the full JT case most likely manifests itself in a highly non-trivial manner, as we saw here for the case of topological gravity. 

Nevertheless, under the conjectured existence of the universal RMT regime, the $\tau-$scaled limit of the SFF of unorientable JT gravity could be computed from the RMT side. This approach would have the advantage of being less technically involved and it can be solved to all orders in $\tau$. In this context, it is interesting to ask whether or not the infinities of the unorientable WP volumes would survive the $\tau-$scaled limit of the SFF, an aspect that we leave for future investigation. Inverting the reasoning, now from the perspective of computing the canonical SFF on the RMT side, i.e.~\eqref{eq:SFF1}, it is not clear where the divergences would arise. This would imply that any $\epsilon$-dependent regularization scheme of the unorientable WP volumes, such as that in \cite{Stanford2023} used to compute the SFF, would have to be independent of $\epsilon$ in the $\tau-$scaled limit. In this way, by assuming agreement with the finite results of universal RMT, well motivated by our results for topological gravity, late-time correlation functions can be computed in unorientable JT gravity. The computation of the orthogonal SFF from universal RMT using the JT gravity leading order energy density will be the subject of future work. 

Furthermore, the computation can also be extended to a matrix model in the symplectic symmetry class since the form factor for this case is known. It is also rather straightforward to generalize the results of the unorientable Airy WP volumes to the symplectic case \cite{Stanford2019}. Along these lines, it would be interesting to see the work extended to the $(\alpha, \beta)$ ensembles, which in the case of $(\alpha,\beta)=(1,2)$ was already done in \cite{Griguolo2024}.

As a further development of the work presented here, the examples of cancellations discussed in \cref{sec:Cancellations} can be extended to the complete set of cancellations as discussed for the orientable case in \cite{Weber2022,Blommaert2022}. A prime application of such a complete set of constraints would be to study whether the equivalence of cancellations to certain properties of intersection numbers found in \cite{Blommaert2022} can be generalised to the unorientable setting. However, this requires a notion of intersection numbers for the unorientable setting, the mathmatically rigorous definition of which as far as we know is an unsolved problem. Along this line of thought it is also tempting to conjecture that the relation between the ``unorientable intersection numbers'' and the unorientable Airy WP volumes is a direct generalisation of what happens in the unitary case (see e.g. \cref{eq:WP_Intersection}), i.e. the intersection numbers correspond to the coefficients of the Airy WP volumes up to some prefactors. This would allow one to investigate whether properties of the orientable intersection numbers, such as the string and dilaton equations, carry over to the unorientable setting. 

Another interesting perspective comes from our observation of logarithmic terms in the canonical SFF for the unorientable Airy model, in contrast to the observation of the cancellation of logarithmic contributions to the canonical SFF of the orientable model in \cite{Saad2022}. In the work for the orientable case, i.e. dual to a unitary matrix model, these cancellations were linked to the cancellations of contributions from certain encounters arising in the periodic-orbit approach to universal RMT for the unitary symmetry class \cite{Muller2005}. This connection was motivated by the study of Kontsevich graphs (cf. \cref{sec:Diagrammatics}). In the periodic-orbit calculation for the orthogonal symmetry class these cancellations are not present, which is exactly what we observe for the logarithmic contributions to the canonical SFF of the unorientable Airy model corresponding to this symmetry class. Thus, it is a well motivated idea to assume that the link between logarithmic contributions and encounters also persists in the orthogonal case, which would then manifest itself in the presence of logarithmic terms in the canonical SFF. To further support this, it would be interesting to consider the explicit form of the unorientable Kontsevich diagrams for special cases such as $g=1$ in order to study more closely the connection between encounter contributions and Kontsevich diagrams in the orthogonal symmetry class.

\acknowledgments
We thank M. Lents for valuable discussions and comments on the manuscript. J.T. would like to thank S. Tomsovic for helpful discussions. We acknowledge financial support from the Deutsche Forschungsgemeinschaft (German Research Foundation) through Ri681/15-1 (project number 456449460) within the Reinhart-Koselleck Programme.
\newpage

\appendix

\section{\boldmath Collection of resolvents for the Airy model, \texorpdfstring{$\upbeta=1$}{\upbeta=1}}\label{app:Resolvents}
To abbreviate the presentation in the following, we introduce the notation
\begin{align}
    m_{a_1,\cdots, a_n}(z_1,\cdots,z_n):=\sum_{\pi\in S(n)}\prod_{i=1}^{n}z_i^{a_{\pi(i)}},
\end{align}
where $S(n)$ denotes the permutation group for a set of order $n$. Note, that here, in contrast to e.g. \cite{Weber2022} and the following \cref{app:Volumes} there is no additional factor of 2 in the exponents.
\begin{align}
    &R_{0}\left(z_1,z_2\right)=\frac{1}{2 z_1 z_2 \left(z_1+z_2\right){}^2}\\
    &R_{\frac 12}\qty(z_1,z_2)=\frac{m_4+3 m_{1,3}+3m_{2,2}}{2 z_1^4 z_2^4 \left(z_1+z_2\right){}^3}\\
    &R_1\qty(z_1,z_2)=\frac{35 m_8+140 m_{1,7}+231 m_{2,6}+240 m_{3,5}+240 m_{4,4}}{16 z_1^7 z_2^7 \left(z_1+z_2\right){}^4}\\
    &\begin{aligned}
        R_{\frac 32}\qty(z_1,z_2)=\frac{1}{16 z_1^{16}z_2^{16}\left(z_1+z_2\right){}^5} \Big[&256 m_{12}+1280 m_{1,11}+2752 m_{2,10}+3590 m_{3,9}\\
        &+3710 m_{4,8}+3739 m_{4,7}+3750 m_{6,6}\Big]
    \end{aligned}
\end{align}
\begin{align}
    &\begin{aligned}
        R_2\qty(z_1,z_2)=\frac{1}{256 z_1^{13} z_2^{13} \left(z_1+z_2\right){}^6}\Big[&42735 m_{16}+256410 m_{1,15}+675990 m_{2,14}+\\
        &+1072682 m_{3,13}+1245767 m_{3,12}+1274496 m_{5,11}+\\
        &+1283626 m_{6,10}+1288828 m_{7,9}+1290516 m_{8,8}\Big]
    \end{aligned}\\
    &\begin{aligned}
        R_{\frac 52}\qty(z_1,z_2)=\frac{1}{256 z_1^{16} z_2^{16} \left(z_1+z_2\right){}^7}\Big[&573440 m_{20}+4014080 m_{1,19}+12533760 m_{2,18}+\\
        &+23596510 m_{3,17}+31441170 m_{4,16}+34299125 m_{5,15}+\\
        &+34838419 m_{6,14}+35034993 m_{7,13}+35143367 m_{8,12}+\\
        &+35195056 m_{9,11}+35210000 m_{10,1}\Big]
    \end{aligned}\\
    &\begin{aligned}
        R_3\qty(z_1,z_2)=\frac{5}{2048 z_1^{19} z_2^{19} \left(z_1+z_2\right){}^8}\Big[&15094079 m_{24}+120752632 m_{1,23}+435952517 m_{2,22}+\\
        &+953649872 m_{3,21}+1456609553 m_{4,20}+1744284376 m_{5,19}+\\
        &+1836576352 m_{6,18}+1855963688 m_{7,17}+1863669631 m_{8,16}+\\
        &+1867877552 m_{9,15}+1870061755 m_{10,14}+1871078024 m_{11,13}+\\
        &+1871370530 m_{12,12}\Big ]
    \end{aligned}
\end{align}
\section{Collection of the unorientable Airy WP volumes}\label{app:Volumes}
Presenting the volumes in the form given in \cref{eq:AiryWP_form_conjecture} becomes rather cumbersome for increasing genus. Instead, we present them in a more symmetric form, i.e.
\begin{align}\label{eq:V_unor}
V^{\text{Airy}}_{g,2}=P_g^1\qty(b_1,b_2)+\theta\qty(b_1-b_2)P_g^2(b_1,b_2)+\theta\qty(b_2-b_1)P_g^2(b_2,b_1),
\end{align}
where $P_g^1$ is the symmetric part of the volume, defined in the following way: For each unordered pair of indices $(n,m)$ appearing in the respective $V^>_g$ find the terms $ C_{m,n}b_1^mb_2^n+C_{n,m}b_1^nb_2^m\subset V_g^>(b_1,b_2)$. Now define $\Tilde{C}_{m,n}=\min(C_{m,n},C_{n,m})$. Then 
\begin{equation}
P_g^1(b_1,b_2)\coloneqq\sum_{m,n}\Tilde{C}_{n,m}(b_1^mb_2^n+b_1^nb_2^m),
\end{equation}
$P_g^2(b_1,b_2)$ can then be computed by subtracting $P_g^1$ from $V^>_g$. Having written the unorientable Airy WP volume in this form, one can of course uniquely recover the form of \cref{eq:AiryWP_form_conjecture} by rewriting $P_g^1(b_1,b_2)=P_g^1(b_1,b_2)\qty[\theta(b_1-b_2)+\theta\qty(b_2-b_1)]$.

Using this convention however, the unorientable Airy WP volumes for $n=2$ and the two smallest genera are 
\begin{align}
    V^{\text{Airy}}_{\frac 12,2}(b_1,b_2)&=\max\qty(b_1,b_2)=\theta\qty(b_1-b_2)b_1+\theta\qty(b_2-b_1)b_2\\
    V^{\text{Airy}}_{ 1,2}(b_1,b_2)&=\frac{1}{96} \left(\left(4 b_1^4+8 b_2^3 b_1\right) \theta \left(b_1-b_2\right)+\left(4 b_2^4+8 b_1^3 b_2\right) \theta \left(b_2-b_1\right)+14 b_1^2 b_2^2+3 \left(b_1^4+b_2^4\right)\right).
\end{align}
To abbreviate the presentation of the volumes of higher genus, we introduce the notation of \cite{Do2011}
\begin{align}
    m_{a_1,\cdots, a_n}(L_1,\cdots,L_n):=\sum_{\pi\in S(n)}\prod_{i=1}^{n}L_i^{a_{\pi(i)}},
\end{align}
where $S(n)$ denotes the permutation group for a set of order $n$.
\begin{align}
    &\begin{aligned}
    P_\frac 3 2^1\qty(b_1,b_2)&=\frac{23 m_{7}}{40320}+\frac{7 m_{5,2}}{1920}+\frac{7 m_{4,3}}{1152}\\
    P_\frac 3 2^2\qty(b_1,b_2)&=\frac{41 b_1^7}{40320}+\frac{43 b_2^2 b_1^5}{5760}+\frac{1}{128} b_2^4 b_1^3+\frac{5 b_2^6 b_1}{1152}
    \end{aligned}
    \\
    &\begin{aligned}
        P_2^1\qty(b_1,b_2)&=\frac{377 m_{10}}{46448640}+\frac{863 m_{8,2}}{5160960}+\frac{m_{7,3}}{7560}+\frac{97 m_{6,4}}{122880}+\frac{m_{5,5}}{3600}\\
        P_2^2\qty(b_1,b_2)&=\frac{5 b_1^{10}}{580608}+\frac{3 b_2^2 b_1^8}{35840}+\frac{b_2^4 b_1^6}{69120}+\frac{b_2^7 b_1^3}{15120}+\frac{b_2^9 b_1}{22680}
    \end{aligned}
    \\
    &\begin{aligned}
        P_\frac 5 2 ^1\qty(b_1,b_2)&=\frac{907 m_{13}}{20437401600}+\frac{32743 m_{11,2}}{30656102400}+\frac{37 m_{10,3}}{26542080}+\\
        &+\frac{377 m_{9,4}}{53084160}+\frac{37 m_{8,5}}{5898240}+\frac{143 m_{7,6}}{13271040}\\
        P_\frac 5 2^2\qty(b_1,b_2)&=\frac{46547 b_1^{13}}{797058662400}+\frac{49177 b_2^2 b_1^{11}}{30656102400}+\frac{50527 b_2^4 b_1^9}{5573836800}+\frac{10483 b_2^6 b_1^7}{464486400}+\\
        &+\frac{4351 b_2^8 b_1^5}{265420800}+\frac{551 b_2^{10} b_1^3}{132710400}+\frac{13 b_2^{12} b_1}{29491200}
    \end{aligned}\\
    &\begin{aligned}
        P_3 ^1\qty(b_1,b_2)&=\frac{149189 m_{16}}{714164561510400}+\frac{2668427 m_{14,2}}{267811710566400}+\frac{m_{13,3}}{116756640}+\\
        &+\frac{246733 m_{12,4}}{1961990553600}+\frac{m_{11,5}}{14968800}+\frac{3121 m_{10,6}}{6370099200}+\\
        &+\frac{m_{7,9}}{4762800}+\frac{86683 m_{8,8}}{107017666560}\\
        P_3^2\qty(b_1,b_2)&=\frac{43 b_1^{16}}{209227898880}+\frac{1153 b_2^2 b_1^{14}}{174356582400}+\frac{211 b_2^4 b_1^{12}}{5748019200}+\frac{47 b_2^6 b_1^{10}}{522547200}+\\
        &+\frac{b_2^9 b_1^7}{14288400}+\frac{b_2^{11} b_1^5}{14968800}+\frac{b_2^{13} b_1^3}{58378320}+\frac{b_2^{15} b_1}{583783200}
    \end{aligned}
\end{align}

\section{Unorientable Airy WP volumes: The point of view of Kontsevich graphs}\label{sec:Diagrammatics}
In this section the diagrammatic approach stemming from the Kontsevich matrix integral of the orthogonal symmetry class will be used to derive the generic structure of the unorientable Airy WP volumes. It is well known the Kontsevich matrix integral of the unitary symmetry class can be taken as a generating function for intersection numbers \cite{Kontsevich1992}. The known relationship between intersection numbers and Airy WP volumes allows a direct computation of the Airy WP volumes from the ribbon graph expansion of the matrix model. It is an obvious question to ask whether or not the Kontsevich matrix integral of the orthogonal symmetry class can be used to compute unorientable Airy WP volumes. The idea was first explored for the case of $(g,n)=(\frac{1}{2},2)$ in \cite{Saad2022}, and the graphs were shown to have the necessary structure for reproducing the correct unorientable Airy WP volume. However, for higher order cases the graphs becomes far too tedious to be used to compute the unorientable Airy WP volumes\footnote{It is also possible the graphs generated by the Kontsevich matrix integral of orthogonal symmetry class need an additional structure, such as orientation reversal, in order to correctly reproduce the unorientable Airy WP volumes.} and the loop equation implemented in section \ref{sec:AiryWP} have to be used. Fortunately, for our purposes, it is possible to derive the generic structure of the unorientable Airy WP volumes from the functional form of the graphs without ever needing to draw the graphs or compute their symmetry factors. The advantage of the diagrammatic method here is the functional form of the graphs is known for all genus. In order to present our arguments we first recall the orientable formalism, primarily based on \cite{Do2008}, comment on the extension to the unorientable setting in section \ref{sec:Diag_form}, and then in section \ref{subsec:AiryWP} go on to derive the general structure of the unorientable Airy WP volumes as claimed in the main text. 
\subsection{General structure}\label{sec:Diag_form}
The diagrammatic expansion we refer to here is an expansion in terms of ribbon diagrams/ribbon graphs, i.e. diagrams with the propagators not being lines but double-lines and analogous vertices. For such a diagram one defines the notion of a boundary by labelling sides of the propagators being connected through a vertex by the same label. Doing this for all sides of all propagators induces a partition of the sides into $n$ distinct sets which is denoted as the number of boundaries of the diagram. Additionally, one can compute the Euler characteristic of the diagram as the Euler characteristic of the corresponding single-line graph.  The number of boundaries of course coincides with the number of boundaries the graph has when being considered as a surface. This point of view allows one to express the Euler characteristic of a given graph as $\chi=2-2g-n$ with $n$ the number of boundaries and $g$ the genus of the diagram when considered as a surface.\\
A convenient starting point for presenting the expression of the Airy WP volumes in terms of ribbon graphs for the orientable case is Kontsevich's theorem which can be stated as
\begin{theorem}[Kontsevich\cite{Kontsevich1992}]
Denote by $\Gamma_{g,n}$ the set of ribbon graphs build from 3-valent vertices having n boundaries and genus g. Furthermore, adopt the standard notation for intersection numbers (cf.\cite{Dijkgraaf2018}). Then it holds that
\begin{align*}
    \sum_{\abs{\Vec{\alpha}}=3g-3+n} \ev{\tau_{\alpha_1},\dots,\tau_{\alpha_n}}\prod_{k=1}^n \frac{\qty(2\alpha_k-1)!!}{z_k^{2\alpha_k+1}} =\sum_{\Gamma_{g,n}}\frac{2^{2g-2+n}}{\abs{\textnormal{Aut}(\Gamma_{g,n})}}\prod_{k=1}^{6g-g+3n}\frac{1}{z_{l(k)}+z_{r(k)}}
\end{align*}
where the product on the RHS is over the edges of the individual ribbon graph, thus $l(k)$ denotes the boundary component to which the left edge of he respective edge belongs, $r(k)$ the right edge, respectively.
\end{theorem}
Therefore, on the side of the ribbon diagrams one only needs to consider 3-valent vertices. This implies that the number of propagators (i.e. number of edges for the one-line graph) is given by $\frac 3 2 \# \text{Vertices}$ as from each vertex three half edges originate that have to recombine with another half edge to form an edge. The Euler characteristic of such a ribbon graph is then
$\chi=\#\text{Vertices}-\#\text{Edges}=-\frac 1 2 \# \text{Vertices}$. From the expression of $\chi$ in terms of genus and number of boundaries one can infer that for all graphs in $\Gamma_{g,n}$
\begin{align}
    \# \text{Vertices}=4g+2n-4 \qcomma \# \text{Edges}=6g+3n-6.
\end{align}
On the intersection number side, to relate to the WP volumes one uses the well known result of Mirzakhani\footnote{For $g=1,n=1$ there is an additional factor of 2 in the convention of \cite{Mirzakhani2006} here we adopt the convention of \cite{Saad2019} for which the factor of 2 is put in the definition of the volume such that it does not appear in this equation.} \cite{Mirzakhani2006}
\begin{align}\label{eq:WP_Intersection}
    V_{g,n}(L_1,\dots,L_n)=\sum_{\abs{\vec{\alpha}}+m=3g-3+n}\frac{\qty(2\pi^2)^m}{2^{\abs{\Vec{\alpha}}} \Vec{\alpha}! m!} \int_{\overline{\mathcal{M}}_{g,n}} \psi_1^{\alpha_1}\cdots \psi_n^{\alpha_n} \omega^m \Vec{L}^{2\Vec{\alpha}}
\end{align}
with $\Vec{\alpha}!\coloneqq \prod_i \alpha_i !$, $\Vec{L}^{\Vec{\alpha}}=\sum_i L_i^{\alpha_i}$. For the further appearing objects, the Weil-Petersson form $\omega$, the Deligne-Mumford compactifiction of the moduli space of Riemann surfaces $\overline{\mathcal{M}}_{g,n}$ and the $\psi$ classes we refer to \cite{Mirzakhani2006} as they will not be needed in the following. This is due to the fact that the Airy WP volumes correspond to the leading order terms of the actual WP volumes \cite{Saad2019}. This corresponds exactly to the terms in \cref{eq:WP_Intersection} without insertion of the WP symplectic form $\omega$. Furthermore, those terms are recognized to be precisely the definition of the intersection numbers, i.e. 
\begin{align}
\ev{\tau_{\alpha_1},\dots,\tau_{\alpha_n}}=\int_{\overline{\mathcal{M}}_{g,n}} \psi_1^{\alpha_1}\cdots \psi_n^{\alpha_n}.
\end{align}
Explicitly one finds
\begin{align}
    V^{\text{Airy}}_{g,n}(L_1,\dots,L_n)=\sum_{\abs{\vec{\alpha}}=3g-3+n}\frac{1}{2^{\abs{\Vec{\alpha}}} \Vec{\alpha}!} \ev{\tau_{\alpha_1},\dots,\tau_{\alpha_n}} \Vec{L}^{2\Vec{\alpha}}.
\end{align}
Recognising the LHS of Kontsevich's theorem to be closely related to the Laplace transform of this expression one computes
\begin{align}
\begin{aligned}
    \mathcal{L}\qty[V^{\text{Airy}}_{g,n}(L_1,\dots,L_n);\qty(z_1,\dots,z_n)]&=\sum_{\abs{\vec{\alpha}}=3g-3+n}\frac{1}{2^{\abs{\Vec{\alpha}}} \Vec{\alpha}!} \ev{\tau_{\alpha_1},\dots,\tau_{\alpha_n}} \prod_{i=1}^{n}\frac{\qty(2a_i+1)!}{\qty(z_i^2)^{a_i+1}}\\
    &=\sum_{\abs{\vec{\alpha}}=3g-3+n}\ev{\tau_{\alpha_1},\dots,\tau_{\alpha_n}}\prod_{k=1}^n \frac{\qty(2\alpha_k-1)!!}{z_k^{2\alpha_k+1}}\\
    &=\sum_{\Gamma_{g,n}}\frac{2^{2g-2+n}}{\abs{\text{Aut}(\Gamma_{g,n})}}\prod_{k=1}^{6g-6+3n}\frac{1}{z_{l(k)}+z_{r(k)}}\label{eq:V_Graphs},
\end{aligned}
\end{align}
where Kontsevich's theorem has been used in the last step. By this relation one now has an expression of the Laplace transform of the Airy WP volumes in terms of ribbon graphs and thus a way to compute the volumes.

The discussion so far applied only for the case of orientable ribbon graphs, i.e. the GUE like theory. However, the Kontsevich diagrammatics can be extended to the orthogonal case by allowing propagators to be twisted as it was demonstrated in \cite{Eynard2018, Saad2022}, resulting in contributions from unorientable manifolds. By this reasoning the authors compute $V^{\text{Airy}}_{\frac 12}(b_1,b_2)$, and one could also go on to consider higher genera or other numbers of boundaries. However, one can use loop equations to compute those volumes much more efficiently, and for the explicit results in the main text we used this method. But by \cref{eq:V_Graphs} one has a way of expressing the Laplace transform of the unorientable Airy WP volume at arbitrary genus as a sum over graphs with the contribution of the individual graph being only dependent on the number of the possible propagators occurring in a graph and a rational constant one has to find by counting. Thus, regarding the generic functional form of the volumes, all relevant information is present in this expression. To be more specific, as we are mainly interested in the spectral form factor we can focus on the case of two boundaries. In this case, there are only three possible propagators. First, the two possibilities of having both sides with the same label (denoted as $(11)$ or $(22)$, following the notation in \cite{Saad2022}) and second the possibility of having different labels. As the total number of propagators for a given genus is fixed, it suffices to know the number of two out of three of these options to determine the functional form of the contribution of a specific graph. Choosing the numbers of $(11)$ and $(22)$ propagators, denoted as $n$ and $m$ one can thus write \cref{eq:V_Graphs} for this special case as
\begin{align}
    \mathcal{L}\qty[V^{\text{Airy}}_{g,2}(L_1,L_2);\qty(z_1,z_2)]=\sum_{n,m} C^k_{n,m}\frac{1}{z_1^n z_2^m \qty(z_1+z_2)^{k(g)-n-m}},\label{eq:V_n2_Diag}
\end{align}
with $k(g)=\#\text{Edges}=6g$ and $C^k_{n,m}$ a rational positive constant.
Therefore, while the process of constructing the graphs, computing the symmetry factors, etc., which fixes the constant $C^k_{n,m}$, is quite tedious, the functional dependence on the $z_i$ is known without ever drawing one diagram. Thus, to unravel the structure of the WP volumes deriving from a specific graph we only have to perform a Laplace transform of the corresponding term, which will be the focus of the following section.
\subsection{General contribution to the unorientable Airy WP volumes}\label{subsec:AiryWP}
Using \cref{eq:V_n2_Diag} one directly sees that the generic contributions arising from a diagram with a total number of $k$ propagators of which $n$ are of type $\qty(11)$ and $m$ of type $22$ to the unorientable Airy WP volume is given by
\begin{align}\label{eq:def_matcalV}
    \mathcal{V}^k_{n,m}(b_1,b_2)\coloneqq C^k_{n,m}\int_{\mathcal{C}_1}\frac{\dd{z_1}}{2\pi i}\int_{\mathcal{C}_2}\frac{\dd{z_2}}{2\pi i} \frac{e^{b_1 z_1}e^{b_2 z_2}}{z_1^n z_2^m \qty(z_1+z_2)^{k-n-m}},
\end{align}
where the contours are chosen such that they go from $-i\infty$ to $i \infty$ on the right side of all poles of the integrand.

Before giving the result of the computation of this integral we remark that the case $(n,m)=(0,0)$, i.e. a diagram with all propagators having different labels on their two sides, is not possible.
\begin{figure}[h]
    \centering
    \includegraphics[height =3cm]{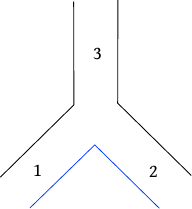}
    \caption{labelled 3-valent vertex}
    \label{fig:Vertex_numberered}
\end{figure}
In order to see this, we look at one 3-valent vertex which is the basic construction block of all diagrams considered above. We choose an ordering of the legs, as indicated by the numbers which is arbitrary. Assuming that one can build a graph having no $(1,1)$/$(2,2)$ propagators, one has to be able to colour the sides of the outgoing propagators\footnote{Being more rigorous one should rather talk about half-edges here, but this is not vital for this argument.} such that every leg has two different colours. We choose to satisfy this, without loss of generality, first for legs 1 and 2, for which there is only one option to do so (modulo interchanging colour labels). Having done so we see that for leg 3, as it not possible to have a line change its colour through a vertex, both sides have the same label. Thus, one cannot even construct a vertex satisfying the property assumed for the whole diagram which contradicts the assumption and thus proves our statement.

Following the same construction one can prove, that there is no graph having no $(1,2)$ propagator as every graph having two boundaries at least contains one vertex of the form depicted in \cref{fig:Vertex_numberered} which necessitates at least one propagator of the form $(1,2)$. Having discussed this, we come to the evaluation of the $\mathcal{V}^k_{n,m}$.\\

As usual, this reduces to the computation of residua at poles for which the following statement, provable by induction (for $\beta>0$), is useful
\begin{align}
    \dv[\alpha]{}{z}\frac{e^{bz}}{\qty(z+a)^\beta}=e^{bz}\sum_{i=0}^\alpha\qty(-1)^{i+\alpha}\frac{\alpha ! \qty(\alpha-i+\beta-1)!}{i!\qty(\alpha-i)!\qty(\beta-1)!}\frac{b^i}{\qty(z+a)^{\alpha+\beta-i}}.\label{eq:Derivative}
\end{align}
Starting here with the $z_2$ integral one can compute, using the residue theorem (one has to close in the half-plane of negative real value of $z_2$ to eliminate the contribution of the closing-contour, thus encircling all the poles) assuming for now $m>0$
\begin{align}
    \int_{\mathcal{C}_2}\frac{\dd{z_2}}{2\pi i} \frac{e^{b_2 z_2}}{z_2^m \qty(z_1+z_2)^{k-n-m}}&=\qty[\Res_{z=0}+\Res_{z=-z_1}]\frac{e^{b_2 z}}{z^m\qty(z_1+z)^{k-n-m}}.
\end{align}
We begin by evaluating the residue at 0:
\begin{align}
\begin{aligned}
    \Res_{z=0}\qty[\frac{e^{b_2 z}}{z^m\qty(z_1+z)^{k-n-m}}]&=\frac{1}{\qty(m-1)!}\eval{\dv[m-1]{}{z}}_{z=0}\frac{e^{b_2 z}}{\qty(z_1+z)^{k-n-m}}\\
    &=\sum_{i=0}^{m-1}\qty(-1)^{i+m-1}\frac{\qty(k-i-n-2)!}{i!\qty(m-1-i)!\qty(k-n-m-1)!}\frac{b_2^i}{\qty(z_1)^{k-n-i-1}},
\end{aligned}
\end{align}
where we used \cref{eq:Derivative} to get to the second line.
Performing now the $z_1$ integral for this contribution one notices that due to the only exponential term being $e^{b_1 z_1}$ one has to close the contour in the half-plane of negative real value of $z_1$, for all values of $b_1\geq 0$. As the sole pole of the integrand is at $z_1=0$ the final contribution is given by
\begin{equation}
\begin{aligned}
    \Res_{z_1=0}&\sum_{i=0}^{m-1}\qty(-1)^{i+m-1}b_2^i\frac{\qty(k-i-n-2)!}{i!\qty(m-1-i)!\qty(k-n-m-1)!}\frac{e^{z_1 b_1}}{\qty(z_1)^{k-i-1}}\\
    &=\sum_{i=0}^{m-1}\qty(-1)^{i+m-1}\frac{\qty(k-i-n-2)!}{i!\qty(m-1-i)!\qty(k-n-m-1)!\qty(k-i-2)!}b_1^{k-i-2} b_2^i\\
    &\coloneqq \mathcal{A}^{k}_{n,m}.
\end{aligned}
\end{equation}
Thus all contributions are of combined order $k-2$ in the lengths.

Coming now to the other residue, one evaluates
\begin{align}
    \begin{aligned}
        \Res_{z=-z_1}&\qty[\frac{e^{b_2 z}}{z^m\qty(z_1+z)^{k-n-m}}]=\frac{1}{(k-n-m-1)!}\eval{\dv[k-n-m-1]{}{z}}_{z=-z_1}\frac{e^{b_2 z}}{z^{m}}\\
        &=e^{-b_2 z_1}\sum_{i=0}^{k-n-m-1}\qty(-1)^{-m}\frac{\qty(k-n-i-2)!}{i!\qty(k-n-m-1-i)!\qty(m-1)!}\frac{b_2^i}{\qty(z_1)^{k-n-i-1}},
    \end{aligned}
\end{align}
again using \cref{eq:Derivative}. Analogously to above the final contribution is evaluated by closing the contour $\mathcal{C}_1$. Here one has to be careful, since the exponential term is now $e^{(b_1-b_2)z_1}$, and one has to close the contour in the half-plane of positive real value of $z_1$ if $b_1-b_2<0$. In this case there is no pole inside the integration region and the integral vanishes. For the case of $b_1-b_2>0$ the pole at $0$ is the only contribution: 
\begin{align}
    \begin{aligned}
        \Res_{z_1=0}&\sum_{i=0}^{k-n-m-1}\qty(-1)^{-m}\frac{\qty(k-n-i-2)!}{i!\qty(k-n-m-1-i)!\qty(m-1)!}\frac{e^{\qty(b_1-b_2) z_1}}{\qty(z_1)^{k-i-1}}b_2^i\\
        &=\sum_{i=0}^{k-n-m-1}\qty(-1)^{-m}\frac{\qty(k-n-i-2)!}{i!\qty(k-n-m-1-i)!\qty(m-1)!\qty(k-i-2)!}\qty(b_1-b_2)^{k-i-2}b_2^i\\
        &\coloneqq \mathcal{B}^{k}_{n,m},
    \end{aligned}
\end{align}
and the contibution to $\mathcal{V}_{n,m}$ arising from this part of the result is given by $\mathcal{B}^{k}_{n,m}\theta(b_1-b_2)$ with $\theta$ denoting the Heaviside function.

Considering briefly the case of either $n$ or $m$ vanishing, (both can't vanish due to the considerations at the beginning of this section) we consider without loss of generality $m=0$. In this case there is no pole at $z_2=0$, and one has to work out only the second residue in the first step i.e. 
\begin{align}
\Res_{z=-z_1}&\qty[\frac{e^{b_2 z}}{\qty(z_1+z)^{k-n}}]=\frac{1}{(k-n-1)!}\eval{\dv[k-n-1]{}{z}}_{z=-z_1}e^{b_2 z}\\
&=\frac{1}{(k-n-1)!}b_2^{k-n-1}e^{b_2 z_1}.
\end{align}
The computation of the second residue is not altered, i.e. one computes
\begin{align}
    \Res_{z_1=0}&\frac{1}{(k-n-1)!}b_2^{k-n-1}\frac{e^{\qty(b_1-b_2)z_1}}{z_1^n}\\
    &=\underbrace{\frac{1}{(k-n-1)!(n-1)!}b_2^{k-n-1}\qty(b_1-b_2)^{n-1}}_{\coloneqq \mathcal{B}^{k}_{n,0}}\theta(b_1-b_2)
\end{align}
As a final result this yields
\begin{align}\label{eq:Airy_WP}
    \mathcal{V}^k_{n,m}(b_1,b_2)=C^k_{n,m}\qty(\mathcal{A}^{k}_{n,m}+\theta(b_1-b_2)\mathcal{B}^{k}_{n,m}).
\end{align}

In all these cases one can easily get the respective $(m,n)$ expression from the $(n,m)$ expression by exchanging $b_1\leftrightarrow b_2$ in the expression and also the $\theta$-function for the case of the $\mathcal{B}$ term. Additionally note, that the coefficients $C^k_{n,m}$ are symmetric upon $n \leftrightarrow m$ as the exchange is just the other labelling of boundaries of the respective graph. By these two observations, one can argue for the case of $n\neq m$ that the contribution to the volume resulting from the diagrams $(n,m)$ and $(m,n)$ is symmetric with respect to $b_1\leftrightarrow b_2$. 

For the case of the contributions from $(n,n)$ type diagrams this is not obvious from \cref{eq:Airy_WP}, as there one cannot make the above argument due to there being (at most) one diagram of this type. However one can see the symmetry with respect to exchanging the boundary lengths in this case by inspecting \cref{eq:def_matcalV}, where it is present. Thus, this has to carry over to the results of the integral so that it has to hold that for all possible $n$ for a given $k$ that,
\begin{align}
    \mathcal{A}^{k}_{n,n}(b_1,b_2) + \mathcal{B}^{k}_{n,n}(b_1,b_2)=\mathcal{A}^{k}_{n,n}(b_2,b_1).
\end{align}
We have observed this to be true for all cases we have checked, and it should follow from the explicit form of $\mathcal{A}^{k}_{n,n}$ and $\mathcal{B}^{k}_{n,n}$. However, as we can also see this by the above argument from the definition we don't go into the tedious work of this proof. Thus, as claimed in the discussion of the WP volumes following from the loop equations we can see the symmetry of the volumes under $b_1\leftrightarrow b_2$ from the diagrammatics.

For the further structure of the WP volumes we can also infer the properties we found in the main text by inspection of the results of the loop equations, i.e. the structure claimed in \cref{eq:AiryWP_form_conjecture}. The form directly follows from \cref{eq:Airy_WP}. To see this we note again that for every diagram of type $(n,m)$ there is a diagram of type $(m,n)$ which, easiest seen by computing its $\mathcal{B}$ contribution by exchanging $b_1\leftrightarrow b_2$ gives a contribution of the type $(\text{Polynomial}\times \theta(b_2-b_1))$. Combining these two diagrams one finds the form we use in \cref{eq:V_unor} to present the volumes which is equivalent to the form of \cref{eq:AiryWP_form_conjecture}. Furthermore we can easily read off that the $V^{>}_g(b_1,b_2)$ are polynomial in $b_1,b_2$ with the combined order of $(k-2)$. Inserting the expression of $k$ in terms of genus, we find the order $6g-2$ which proves the final piece of the structure claimed in \cref{eq:AiryWP_form_conjecture}.

While this concludes the proof of the form of the unorientable Airy WP volumes made in the main text we could go on to proof the structure claimed for the contributions to the genus expansion of the resolvents as well. For the sake of convenience we recall from the main text that the relation of the WP volumes with the resolvents is given by \cref{eq:V[R]}
\begin{align}
    V_{g,n}(b_1,\dots,b_n)=\mathcal{L}^{-1}\qty[\prod_{i=1}^n\qty( \frac{-2z_i}{b_i})R_g(z_1,\dots,z_n),\qty(b_1,\dots,b_n)],
\end{align}
which can conveniently be rewritten as a relation including the Laplace transform of the volumes, i.e. 
\begin{align}\label{eq:R[L_V]}
    \mathcal{L}\qty[\qty(\prod_{i=1}^n b_i)V_{g,n}(b_1,\dots,b_n),\qty(z_1,\dots,z_n)]&=\prod_{i=1}^n\qty(-2z_i)R_g(z_1,\dots,z_n)\\
    &=\qty(-1)^n\prod_{i=1}^n \pdv{z_i}\mathcal{L}\qty[V_{g,n}(b_1,\dots,b_n),\qty(z_1,\dots,z_n)],
\end{align}
where in the second line we used a well-known property of the Laplace transform. As we know the general structure of the Laplace transform of the unorientable Airy WP volumes expressed by diagrammatics, given by \cref{eq:V_n2_Diag}, we can perform the derivative and write the sum with a common denominator, keeping in mind the restriction of the number of propagators of the form $(1,2)$ being in $[1,k-1]$. Doing this, we find
\begin{align}
    R^{\text{d(iagrammatic)}}_g(z_1,z_2)=\frac{P^d_g(z_1,z_2)}{(z_1 z_2)^{6g+1} (z_1+z_2)^{6g}},
\end{align}
with the combined order of the polynomial $P^d_g(z_1,z_2)$ being $12g-2$. As claimed in the main text, this coincides with the functional form of the resolvents given by the loop equations while producing, in general, higher orders for the polynomial $P^d_g$ and in the inverse sum of the variables. Curiously, for genus $\frac 12$ the structures coincide. However, for higher genera this does not hold true, and the coefficients $C^k_{n,m}$ in \cref{eq:V_n2_Diag} have to be such that $P^d_g(z_1,z_2)$ has a zero of order $4g-2$ at $z_1=-z_2$. As we do not consider the specific diagrammatics, i.e. constructing explicitly the diagrams this is a check of the consistency of the diagrammatics with the loop equations that is left for further study.

\section{The canonical SFF: General structure}\label{sec:SFF_general}
The contributions to the topological expansion of the canonical SFF at genus $g$, $\kappa^g_\beta(t)$ can, as stated in the main text, be computed as
\begin{align}
    \kappa^g_\beta(t)=\int_0^\infty \dd{b_1}b_1 \int_0^\infty \dd{b_2}b_2 Z^t(\beta_1,b_1) Z^t(\beta_2,b_2) V_{g,2}(b_1,b_2)
\end{align}
with $Z^t(\beta,b)=\frac{1}{\sqrt{4\pi \beta}}e^{-\frac{b^2}{4\beta}}$ and $\beta_1=\beta+ i t $, $\beta_2=\beta_1^*$. The results of this computations were reported above for the explicit volumes derived from the loop equations. For the main purpose of this work, i.e. the comparison of the SFF with the predictions of universal RMT, this suffices and the explicit computation is of no special interest, except that it follows through without any problems. However, if one wishes to study the relation of the specific terms of the canonical SFF with the various contributions to the unorientable Airy WP volumes or, in the spirit of \cite{Weber2022}, wishes to infer constraints on the coefficients of the unorientable Airy WP volumes it is of interest what the contribution to the SFF for a generic unorientable Airy WP volume is. Thus, in this appendix we compute this, starting from the generic form of the unorientable Airy WP Volume, stated in \cref{eq:AiryWP_form_conjecture}. After having done this, in section \ref{sec:arctan} we will turn our interest to a more detailed treatment of the $\arctan$, arising for specific contributions to the SFF of the unorientable Airy model.
\subsection{Contribution from generic $V^{\text{Airy}}_{g,2}$}
We recall the generic form of the Airy WP Volume in the unorientable case to be 
\begin{align}
    V^{\text{Airy}}_{g,2}(b_1,b_2)=V^>_g(b_1,b_2)\theta\qty(b_1-b_2)+V^>_g(b_2,b_1)\theta\qty(b_2-b_1),
\end{align}
or equivalently
\begin{align}
    V^{\text{Airy}}_{g,2}=P_g^1\qty(b_1,b_2)+\theta\qty(b_1-b_2)P_g^2(b_1,b_2)+\theta\qty(b_2-b_1)P_g^2(b_2,b_1),
\end{align}
with $V^>_g$ and $P_g^i$ being polynomials in the two lengths.

Thus, it actually suffices to consider the contribution to the SFF of terms of the form $b_1^\alpha b_2^{\beta}\theta\qty(b_1-b_2)$. For the sake of completeness however, we also consider the case of purely polynomial terms, arising from $P_g^1$, which can be dealt with easily (almost) analogous to the orientable case, i.e.
\begin{align}
    \begin{aligned}
        &\int_0^\infty\dd{b_1}b_1\int_0^{\infty}\dd{b_2}b_2 \frac{e^{-\frac{b_1^2}{4\beta_1}}}{\sqrt{4\pi\beta_1}}\frac{e^{-\frac{b_2^2}{4\beta_2}}}{\sqrt{4\pi\beta_2}}b_1^a b_2^b\\
        &=\frac{4\beta_1\beta_2}{4\pi\sqrt{\beta_1\beta_2}}2^a\sqrt{\beta_1}^a\int_0^\infty\dd{x_1}e^{-x_1}x_1^{\frac a 2}2^b\sqrt{\beta_2}^b\int_0^\infty\dd{x_2}e^{-x_2}x_2^{\frac b 2}\\
        &=\frac{2^{a+b}\sqrt{\beta_1\beta_2}\beta_1^{\frac{a}{2}}\beta_2^{\frac{b}{2}}}{\pi}\Gamma\qty(\frac b 2+1)\Gamma\qty(\frac a 2 +1)\\
        &\eqqcolon R(a,b)(\beta_1,\beta_2)
    \end{aligned}
\end{align}
As one can clearly see, upon setting $\beta_1=\beta+i t, \beta_2=\beta_1^*$ this yields $\sqrt{t^2+\beta^2}$ multiplied by a polynomial in $t$ and $\beta$ where some terms can be multiplied by $\sqrt{\beta_i}$ in the case of exactly one of $(a,b)$ being odd. In the case of $(a,b)$ both being odd, as the leading square root can be combined with another such term the result is, even before taking the limit of large $t$, a polynomial in $t$ and $\beta$.

The interesting terms are those arising from the part of the volume containing $\theta$ functions. Thus one has to evaluate the integral
\begin{align}
\begin{aligned}
        I(\alpha,\gamma)\coloneqq&\int_0^\infty\dd{b_1}b_1\int_0^{\infty}\dd{b_2}b_2 \frac{e^{-\frac{b_1^2}{4\beta_1}}}{\sqrt{4\pi\beta_1}}\frac{e^{-\frac{b_2^2}{4\beta_2}}}{\sqrt{4\pi\beta_2}}b_1^\alpha b_2^\gamma \theta(b_1-b_2)\\
        &=\frac{1}{4\pi\sqrt{\beta_1\beta_2}}\int_0^\infty\dd{b_1}e^{-\frac{b_1^2}{4\beta_2}}b_1^{\alpha+1}\int_0^{b_1}\dd{b_2}e^{-\frac{b_2^2}{4\beta_2}}b_2^{\gamma+1}.
    \end{aligned}
\end{align}
This suffices as the contributions arising from the $\theta\qty(b_2-b_1)$-term complementary to $b_1^\alpha b_2^\gamma$, i.e. $b_2^\alpha b_1^\gamma$ results in the same integral $I\qty(\alpha,\gamma)$ with $\beta_1\leftrightarrow \beta_2$.

This integral is dealt with in two ways. First, by using differentiation under the integral sign, which will facilitate the identification of the dependence of the resulting contributions to the SFF as functions of $\beta_1$ and $\beta_2$. With the notation $a\coloneqq \frac{1}{4\beta_2}$ $b\coloneqq \frac{1}{4\beta_1}$ one finds, distinguishing four cases.
\begin{align}
        \gamma \text{ even,}\, \alpha \text{ even} \quad I(\alpha,\gamma)&=\frac{1}{16\pi\sqrt{\beta_1\beta_2}}\qty(-\pdv{}{a})^{\frac \gamma 2}\qty(-\pdv{}{b})^{\frac \alpha 2}\frac{1}{b\qty(a+b)}\\
        \gamma \text{ even,}\, \alpha \text{ odd} \quad I(\alpha,\gamma)&=\frac{1}{16\sqrt{\pi}\sqrt{\beta_1\beta_2}}\qty(-\pdv{}{a})^{\frac {\gamma} 2}\qty(-\pdv{}{b})^{\frac {\alpha+1} 2}\frac{1}{a}\qty[\frac{1}{\sqrt{b}}-\frac 1 {\sqrt{a+b}}]\\
        \gamma \text{ odd,}\, \alpha \text{ even} \quad I(\alpha,\gamma)&=\frac{1}{16\sqrt{\pi}\sqrt{\beta_1\beta_2}}\qty(-\pdv{}{a})^{\frac {\gamma+1} 2}\qty(-\pdv{}{b})^{\frac {\alpha} 2}\frac{1}{b\sqrt{a+b}}\\
        \gamma \text{ odd,}\, \alpha \text{ odd} \quad I(\alpha,\gamma)&=\frac{1}{8\pi\sqrt{\beta_1\beta_2}}\qty(-\pdv{}{a})^{\frac {\gamma+1} 2}\qty(-\pdv{}{b})^{\frac {\alpha+1} 2}\frac{\arctan\qty(\sqrt{\frac a b})}{\sqrt{ab}}.\label{eq:I_oddodd}
\end{align}
Observing those results, one can distinguish three cases for the terms following the common $\frac{1}{\sqrt{\beta_1 \beta_2}}$ term
\begin{enumerate}[i)]
    \item $\gamma,\alpha$ both even: ``standard'' contribution, i.e. purely polynomial in $\beta_1,\beta_2$
    \item $\gamma,\alpha$ containing one even and one odd number: polynomial terms that can be multiplied by $\sqrt{\beta_i}$ and $\sqrt{\beta_1+\beta_2}$
    \item $\gamma,\alpha$ both odd: terms of the type ii) with additional terms of a polynomial multiplying $\arctan\qty(\frac{\beta_1}{\beta_2})$
\end{enumerate}
Thus, as claimed in the main text, we have shown that the appearance of an $\arctan$, which leads to a logarithm in the large $t$ limit as we will see below, is only possible for a term in $V^>_{g,2}$ that is of odd order in both $b_1$ and $b_2$. Furthermore, apart from this type of contribution, one cannot make the direct connection of a specific term appearing in the SFF to a certain type of contribution to the unorientable Airy WP volume.

We now turn to the second way we considered of solving the integral which will result in an expression of $I(\alpha,\gamma)$ as a finite sum, rather than a derivative, which will be most useful when attempting to study other constraints than the ones studied in \cref{sec:Cancellations}. Specifically we use Hypergeometric functions to find for even $\alpha$
\begin{align}
   I(\alpha,\gamma)&=\frac{2^{\alpha +\gamma+1} \beta_1^{\frac{1}{2} (\alpha +\gamma+3)} \beta_2^{\frac{1}{2} (\alpha +\gamma+1)}  \Gamma \left(\frac{1}{2} (\alpha +\gamma+4)\right) }{\pi  (\gamma+2)\left(\beta_1+\beta_2\right){}^{\left(\frac{\alpha +\gamma}{2}+1\right)}}\sum_{k=0}^{\frac{\alpha}{2}} \frac{\frac{\alpha}{2}! \left(\frac{\beta_1}{\beta_2}\right){}^k \Gamma \left(\frac{\gamma}{2}+2\right)}{\left(\frac{\alpha}{2}-k\right)! \Gamma \left(k+\frac{\gamma}{2}+2\right)}.
\end{align}
This also yields the case of odd $\alpha$, even $\gamma$ as one can use $1=\theta(b_1-b_2)+\theta(b_2-b_1)$ to write
\begin{align}
    I(\alpha,\gamma)(\beta_1,\beta_2)=-\tilde{I}(\alpha,\gamma)(\beta_1,\beta_2)+R(\alpha,\gamma)(\beta_1,\beta_2)=-I(\gamma,\alpha)(\beta_2,\beta_1)+R(\alpha,\gamma)(\beta_1,\beta_2),
\end{align}
where $\tilde{I}(\alpha,\gamma)$ denotes the analogue of $I(\alpha,\gamma)$ with $\theta(b_1-b_2)$ replaced by $\theta(b_2-b_1)$.

Thus, we only have to consider the case of both being odd separately. Setting $\alpha=2a+1,\gamma=2b+1$ we find
\begin{align}
    I(\alpha,\gamma)=\frac{2^{2 (a+b)+3} \beta_1^{a+b+\frac{5}{2}} \beta_2^{a+b+\frac{3}{2}}  \Gamma (a+b+3)}{\pi  (2 b+3)\left(\beta_1+\beta_2\right){}^{a+b+2}}\, _2F_1\left(1,-a-\frac{1}{2};b+\frac{5}{2};-\frac{\beta_1}{\beta_2}\right),
\end{align}
where the Hypergeometric function can explicitly be written as
\begin{align}
\begin{aligned}
    _2F_1&\left(1,-a-\frac{1}{2};b+\frac{5}{2};-\frac{\beta _1}{\beta _2}\right)=\\
    &\frac{2 \Gamma \left(a+\frac{3}{2}\right) \Gamma \left(b+\frac{5}{2}\right)}{\pi } \sum _{j=-a-1}^{b+1} \frac{\left(\frac{\beta _1}{\beta _2}\right){}^{-j-\frac{1}{2}}}{(a+j+1)! (b-j+1)!}\\
    &\left(-\sum _{k=0}^{j-1}\frac{(-1)^{k} \sqrt{\frac{\beta _1}{\beta _2}}{}^{2 k+1}}{2 k+1}+\sum _{k=j}^{-1}\frac{(-1)^{k} \sqrt{\frac{\beta _1}{\beta _2}}{}^{2 k+1}}{2 k+1}+\arctan\left(\sqrt{\frac{\beta _1}{\beta _2}}\right)\right),
\end{aligned}
\end{align}
where we adopt the convention that sums where the upper limit is smaller than the lower limit vanish. These results, which are actually quite tedious to obtain, can now be used to write out the contribution to the SFF arising from a generic unorientable Airy WP volume and find cancellations by the method demonstrated in \cref{sec:Cancellations} where it was actually more economical to use the first way of evaluating the integral.

Having finished this computation, we are only left with having to consider the large $t$ expansion of the $\arctan$, appearing in both solutions for $I(\alpha,\gamma)$ to arrive at the form of the SFF which is comparable to the RMT prediction for the canonical SFF. 
\subsection{Treatment of the $\arctan$}\label{sec:arctan}
As we showed in the previous section, the only term arising from the computation of the canonical spectral form factor that doesn't have the ``standard'' functional dependence is the term containing an $\arctan$, arising from the $\theta$ function contribution to the volume with $\alpha,\gamma$ both odd. The aim of this section is to perform carefully the large $t$ expansion for those terms, that was already briefly explained in the main text.

In order to do this, we first determine which terms containing an $\arctan$ can actually appear from $I\qty(\alpha,\gamma)$ with both arguments being odd. As one can see best from the expression for $I\qty(\alpha,\gamma)$ in this case given in \cref{eq:I_oddodd}, the relevant terms are the ones where the derivative doesn't act on the $\arctan$ as this would give a rational function that is not of interest for the present purpose. Thus, the relevant contribution arising from $I_{\alpha,\gamma}$ is given by
\begin{align}
\begin{aligned}
     I^{\log}_{\alpha,\gamma}(\beta_1,\beta_2)\coloneqq \frac{2^{6g-1}\Gamma\qty(1+\frac \alpha 2)\Gamma\qty(1+\frac\gamma 2)}{\pi^2}\beta_1^{\frac{\alpha+1}{2}}\beta_2^{\frac{\gamma+1}{2}}\arctan\qty(\sqrt{\frac{\beta_1}{\beta_2}}).
\end{aligned}
\end{align}
Now we use the choice of $\beta_1$ and $\beta_2$ leading to the SFF, i.e. $\beta_1=\beta+ it=\beta_2^*$.
To simplify further it is useful to use the identity 
\begin{align}
    \arctan{z}=\frac{1}{2i}\log\qty(\frac{1+iz}{1-iz}),
\end{align}
valid for $z\in \mathbb{C}$. Using $z_\pm=\frac{\sqrt{\beta\pm i t}}{\sqrt{\beta\mp i t}}$, thus considering also the case appearing from the complementary term of the unorientable Airy WP volume, i.e. $b_2^\alpha b_1^\gamma \theta(b_2-b_1)$, one finds
\begin{align}
    \arctan\qty(z_\pm)=\frac{1}{2i}\log\qty(i\frac{\mp t +\sqrt{t^2+\beta^2}}{\beta})=\frac{1}{2i}\qty[\log\qty(\frac{\mp t +\sqrt{t^2+\beta^2}}{\beta})+i\frac \pi 2].
\end{align}

Now the task is to expand this expression for large $\frac t \beta \coloneqq \frac 1 x$, i.e. small $x$. Writing the above logarithm in this form, starting with the $+$ case one finds
\begin{align}
\begin{aligned}
    \log\qty(-\frac 1 x +\sqrt{1+\frac{1}{x^2}})&=\log\qty(-\frac 1 x +\sum_{k=0}^{\infty}\binom{\frac 12}{k}x^{2k-1})\\
    &=\log\qty[\frac x 2 \qty(1+2\sum_{k=2}^{\infty}\binom{\frac 12}{k}x^{2k-2})]\\
    &=\log\qty(\frac x 2)+\log\qty(1+2\sum_{k=2}^{\infty}\binom{\frac 12}{k}x^{2k-2})\\
    &=\log\qty(\frac x 2)+\sum_{n=1}^\infty \frac{\qty(-1)^{n+1}}{n}2^n\qty(\sum_{k=2}^{\infty}\binom{\frac 12}{k}x^{2k-2})^n
\end{aligned}
\end{align}
Where in the first line we explicitly use the considered limit of small $x$, more specifically $\frac{1}{x^2}>1$ to use the generalized binomial series. Furthermore, in the last line we used that the absolute value of the sum, for $x<\infty$ is bounded above by $1$ as one can see by resumming the series, which enables one then to use the Mercator series for the logarithm.

The final result, as one can see directly, takes the form of $\log\qty(\frac{\beta}{2t})$ with corrections in even powers of $x$. The coefficients of the corrections terms are elementary, though tediously, computable by using the multinomial theorem for the power of the sum and subsequently collecting powers of $x$. Explicitly one can write
\begin{align}
\log\qty(\frac{-t +\sqrt{t^2+\beta^2}}{\beta})=\log\qty(\frac{\beta}{2t})+\sum_{m=2,4,\dots}C(m)\qty(\frac{\beta}{t})^{m},
\end{align}
with $C(2)=-\frac 14 $, $C(4)=\frac{3}{32}$, $C(6)=-\frac{5}{96}$, $\dots$.\\
Performing the same steps for the complementary case one finds 
\begin{align}
    \log\qty(\frac{t +\sqrt{t^2+\beta^2}}{\beta})=-\log\qty(\frac{\beta}{2t})-\sum_{m=2,4,\dots}C(m)\qty(\frac{\beta}{t})^{m},\label{eq:arctan_comp}
\end{align}
This is all there has to be done for the expansion of the logarithmic part of the $\arctan$. To get the final contributions to the canonical spectral form factor, apart from prefactors, one now only has to include the factors of $\beta_1$ and $\beta_2$. For this we use
\begin{align}
    \beta_1^{\frac{\alpha+1}{2}}\beta_2^{\frac{\gamma+1}{2}}=\sum_{n=}^{\frac{\alpha+1}{2}}\sum_{m=0}^{\frac{\gamma+1}{2}}\binom{\frac{\alpha+1}{2}}{n}\binom{\frac{\gamma+1}{2}}{m}\beta^{\frac{\alpha+\gamma}{2}+1-n-m} \qty(it)^{n+m} \qty(-1)^{m}
\end{align}
which is always a polynomial in $t$ and $\beta$ with a fixed combined order. The notable thing about these polynomials is, as one can easily convince oneself, that terms that have an odd power of $t$ are always accompanied by a factor of $i$ while the even powers are not. Thus, only two cases of terms can appear in $I^{\log}_{\alpha,\gamma}(\beta_1,\beta_2)$. The first of which is for odd powers of $t$  
\begin{align}
\begin{aligned}
     i t^{\text{odd}}\beta^k \arctan(z_+) \rightarrow \frac 12 \Bigg[&t^{\text{odd}}\beta^k \qty(\log\qty(\frac{\beta}{2t})+\sum_{m=2,4,\dots}C(m)\qty(\frac{\beta}{t})^{m} + i \frac\pi 2) \\ -&t^{\text{odd}}\beta^k \qty(-\log\qty(\frac{\beta}{2t})-\sum_{m=2,4,\dots}C(m)\qty(\frac{\beta}{t})^{m}+i\frac{\pi}2)\Bigg ]\\
     &=t^{\text{odd}}\beta^k\qty(\log\qty(\frac{\beta}{2t})+\sum_{m=2,4,\dots}C(m)\qty(\frac{\beta}{t})^{m}),
\end{aligned}
\end{align}
where by $\rightarrow$ we indicate that a term of this form arising in $I^{\log}_{\alpha,\gamma}(\beta_1,\beta_2)$ will result in the rhs of the arrow upon combination with the complementary contribution, for which in the prefactors $t\to-t$ and for the $\arctan$ the complementary result \cref{eq:arctan_comp} is used. Second we consider even powers of $t$ 
\begin{align}
    \begin{aligned}
        t^{\text{even}}\beta^k \arctan(z_+) \rightarrow &\frac{t^{\text{even}}\beta^k}{2 i}\Bigg[\log\qty(\frac{\beta}{2t})+\sum_{m=2,4,\dots}C(m)\qty(\frac{\beta}{t})^{m} + i \frac\pi 2\\
        &~-\log\qty(\frac{\beta}{2t})-\sum_{m=2,4,\dots}C(m)\qty(\frac{\beta}{t})^{m} + i \frac\pi 2\Bigg]\\
        &=\frac{t^{\text{even}}\beta^k \pi}{2}.
    \end{aligned}
\end{align}
This concludes the steps needed to perform the expansion. The reason why this is stated in such detail is the important application of these results for the considerations of cancellation in \cref{sec:Cancellations}.

\section{Derivation of the canonical SFF from universal RMT}
\label{rmt_derive}
In this section the solution to the canonical SFF from universal RMT,~\eqref{eq:SFF_GOE}, will be derived. The integral that needs to be solved is:
\begin{equation}
    \begin{aligned}
        \kappa^{\text{GOE}}_{\beta}(\tau) =
        & e^{S_0}\int_{0}^{\tau^2} \dd{E} e^{-2\beta E}\qty[\frac{\sqrt{E}}{\pi}-\frac{\tau}{2\pi}\log\qty(\frac{\tau}{\pi}+\frac{\sqrt{E}}{2\pi})+\frac{\tau}{2\pi}\log(\frac{\tau}{\pi}-\frac{\sqrt{E}}{2\pi})]+ \\[10pt]
        & e^{S_0}\int_{\tau^2}^{\infty}\dd{E}e^{-2\beta E}\qty[\frac{\tau}{\pi}-\frac{\tau}{2\pi}\log\qty(1+\frac{2\tau}{\sqrt{E}})].
    \end{aligned}
\end{equation}
It can be conveniently written as:
\begin{align}
     &=2\kappa^{\text{GUE}}_\beta(\tau)-e^{S_0}\frac{\tau}{2\pi}\qty[\int_{0}^{\infty} \dd{E} e^{-2\beta E}\log\qty(1+\frac{2\tau}{\sqrt{E}})-\int_{0}^{\tau^2}\dd{E}e^{-2\beta E}\log\qty(-1+\frac{2\tau}{\sqrt{E}})]\\
    &= 2\kappa^{\text{GUE}}_\beta(\tau)+\chi_\beta(\tau)
\end{align}
where 
\begin{align}
    2\kappa^{\text{GUE}}_{\beta}(\tau) = \frac{e^{S_0}}{2\left(2\beta\right)^{3/2}\sqrt{\pi}}\Erf\left(\sqrt{2\beta \tau^2}\right)
    \label{eq:SFF_GUE}
\end{align}
which was first computed in \cite{Saad2022}, and 
\begin{align}
    \chi_{\beta}(\tau)=-e^{S_0}\frac{\tau}{2\pi}\qty[\int_{0}^{\infty} \dd{E} e^{-2\beta E}\log\qty(1+\frac{2\tau}{\sqrt{E}})-\int_{0}^{\tau^2}\dd{E}e^{-2\beta E}\log\qty(-1+\frac{2\tau}{\sqrt{E}})]
\end{align}
Focusing on the bracketed part, the logs can be split up:
\begin{equation}
    \qty[\int_0^{\infty}\dd{E} e^{-2 \beta E} \log\qty(2\tau+ \sqrt{E})-\int_0^{\tau^2} e^{-2\beta E}\log\qty(2\tau - \sqrt{E}) - \int_{\tau^2}^{\infty} e^{-2\beta E}\log\qty(\sqrt{E})]
\end{equation}
Each integral can be integrated by parts and the sum of the boundary terms vanishes.
The remaining integrals are then:
\begin{equation}
    \frac{1}{2\beta}\qty[\int_0^{\infty}\dd{E} e^{-2 \beta E} \frac{1}{2\sqrt{E}(2\tau+ \sqrt{E})}+\int_0^{\tau^2} e^{-2\beta E}\frac{1}{2\sqrt{E}(2\tau - \sqrt{E})} - \int_{\tau^2}^{\infty} e^{-2\beta E}\frac{1}{(2{E})}]
\end{equation}
This can be written as:
\begin{equation}
    \frac{1}{2\beta}\qty[\int_0^{\infty}\dd{x}( e^{-2 \beta (x-\tau)^2}\frac{1}{(x+\tau)} -e^{-2 \beta (x+\tau)^2}\frac{1}{(x+\tau)})]
    \label{eq:J8}
\end{equation}
The translation operator can be used the on the first term:
\begin{equation}
e^{-2\beta(x-\tau)^2} = e^{-2\tau\frac{d}{dx}}e^{-2\beta(x+\tau)^2} = \sum_{n=0}\frac{(-2\tau)^n}{n!}\frac{d^n}{dx^n}e^{-2\beta(x+\tau)^2} 
\end{equation}
The $n=0$ term will cancel the second integral in~\eqref{eq:J8}, so we will start the sum at $n=1$. The integrals in~\eqref{eq:J8} can then be simplified to: 
\begin{equation}
\frac{1}{2\beta}\sum_{n=1}\frac{(-2\tau)^n}{n!}\int_0^{\infty}dx\frac{\frac{d^n}{dx^n}e^{-2\beta(x+\tau)^2}}{(x+\tau)} 
\end{equation}
Making a change a variables $\sqrt{2\beta}(x+\tau) \rightarrow x$ and inserting the identity:
\begin{equation}
\frac{1}{2\beta}\sum_{n=1}\frac{(2\sqrt{2\beta}\tau)^n}{n!}\int_{\sqrt{2\beta}\tau}^{\infty}dx\frac{(-1)^ne^{-x^2}e^{x^2}\frac{d^n}{dx^n}e^{-(x)^2}}{x} 
\label{eq:J11}
\end{equation}
The integrand in~\eqref{eq:J11} is just the Hermite polynomial as defined by:
\begin{equation}
    H_n(x) = (-1)^ne^{x^2}\frac{d^n}{dx^n}e^{-(x)^2} = (n!)\sum_{l=0}^{\text{floor}(n/2)} \frac{(-1)^l(2)^{n-2l}}{(l)!(n-2l)!}(x)^{n-2l}
\end{equation}
This allows a change of variables to be made $x^2 \rightarrow x$. Looking only at the integral:
\begin{equation}
\int_{2\beta\tau^2}^{\infty}dx \frac{1}{2}e^{-x} (x)^{\frac{n}{2} - l -1} = \frac{\Gamma(\frac{n}{2} - l, 2\beta \tau^2)}{2}
\end{equation}
The full solution is then a sum over incomplete gamma functions:
\begin{equation}
\chi_{\beta}(\tau) = -e^{S_0}\frac{\tau}{2\pi}\Bigg{[}\frac{1}{4\beta}\sum_{n=1}^{\infty}\sum_{l=0}^{\text{floor}(n/2)} \frac{(-1)^l(2)^{n-2l}}{(l)!(n-2l)!}(2\sqrt{2\beta}\tau)^n\Gamma(\frac{n}{2} - l, 2\beta \tau^2)\Bigg{]}
\end{equation}
To solve this expression it is necessary to first remove the $\Gamma(0, 2\beta \tau^2)$ from the sum. This occurs only for even $n$. The sum can be split into odd and even terms to remove $l = n/2$ term:
\begin{equation}
\begin{aligned}
\chi_{\beta}(\tau) =&{} -e^{S_0}\frac{1}{4\beta}\frac{\tau}{2\pi}\Bigg{[}  \Gamma(0, 2\beta \tau^2)(e^{-8\beta\tau^2} - 1) + \sum_{n=1}^{\infty}\sum_{l=0}^{n-1} \frac{(-1)^l(2)^{2n-2l}}{(l)!(2n-2l)!}(2\sqrt{2\beta}\tau)^{2n}\Gamma(n - l, 2\beta \tau^2) + \\[.5pt]
& \sum_{n=1}^{\infty}\sum_{l=0}^{n-1} \frac{(-1)^l(2)^{2n-1-2l}}{(l)!(2n-1-2l)!}(2\sqrt{2\beta}\tau)^{2n-1}\Gamma(n - \frac{1}{2} - l, 2\beta \tau^2)\Bigg{]}
\end{aligned}
\end{equation}
Interchanging the sums gives:
\begin{equation}
\begin{aligned}
&{} \sum_{l=0}^{\infty}\sum_{n=l+1}^{\infty} \frac{(-1)^l(2)^{2n-2l}}{(l)!(2n-2l)!}(2\sqrt{2\beta}\tau)^{2n}\Gamma(n - l, 2\beta \tau^2) + \\[.5pt]
& \sum_{l=0}^{\infty}\sum_{n=l+1}^{\infty} \frac{(-1)^l(2)^{2n-1-2l}}{(l)!(2n-1-2l)!}(2\sqrt{2\beta}\tau)^{2n-1}\Gamma(n - \frac{1}{2} - l, 2\beta \tau^2)
\end{aligned}
\end{equation}
A change of variables can be made: $m = n - l$. The two sums can be recombined which gives:
\begin{equation}
\sum_{l=0}^{\infty}\sum_{m=1}^{\infty} \frac{(-1)^l(2)^{m}}{(l)!(m)!}(2\sqrt{2\beta}\tau)^{m+2l}\Gamma(\frac{m}{2}, 2\beta \tau^2) = e^{-8\beta\tau^2}\sum_{m=1}^{\infty}2^m\frac{(2\sqrt{2\beta}\tau)^m}{(m!)}\Gamma(\frac{m}{2}, 2\beta \tau^2)
\end{equation}
where the sum over $l$ was performed. For $m>0$ the incomplete gamma function can be written as:
\begin{equation}
\Gamma(m,x) = \Gamma(m) - \sum_{n=0}^{\infty}\frac{(-1)^n x^{m+n}}{n!(m+n)}
\end{equation}
Plugging this in gives:
\begin{equation}
e^{-8\beta\tau^2}\Bigg{[}\sum_{m=1}^{\infty}2^m\frac{(2\sqrt{2\beta}\tau)^m}{(m!)}\Gamma(\frac{m}{2}) - \sum_{m=1}^{\infty}\sum_{n=0}^{\infty}\frac{2^{2m}(\sqrt{2\beta}\tau)^{2m+2n} (-1)^{n}}{(m!)(n!)(\frac{m}{2}+n)}\Bigg{]} 
\end{equation}
Concentrating on the second term since the first is already in an acceptable form, a change of variable can be made $n + m \rightarrow n$ and then the sums can be interchanged:
\begin{equation}
\sum_{m=1}^{\infty}\sum_{n=0}^{\infty}\frac{2^{2m}(\sqrt{2\beta}\tau)^{2m+2n} (-1)^{n}}{(m!)(n!)(\frac{m}{2}+n)} = \sum_{n=1}^{\infty}\sum_{m=1}^{n}\frac{2^{2m}(\sqrt{2\beta}\tau)^{2n} (-1)^{n+m}}{(m)!(n-m)!(n-\frac{m}{2})}
\end{equation}
The important point is that after interchanging the sums the inner sum becomes finite. This allows simple computation of the coefficients of $2\beta \tau^2$. The full expression is then:
\begin{equation}
\begin{aligned}\label{eq:SFF_GOE_presum}
\chi_{\beta}(\tau) &{}= -e^{S_0}\frac{1}{4\beta}\frac{\tau}{2\pi}\Bigg{[}  \Gamma(0, 2\beta \tau^2)(e^{-8\beta\tau^2} - 1) + \\[.5pt]
& e^{-8\beta \tau^2}\sum_{n=1}^{\infty}\Bigg{(}\frac{(2)^n(2\sqrt{2\beta}\tau)^{n}}{(n)!}\Gamma(\frac{n}{2}) - \sum_{m=1}^{n}\frac{(-1)^{n+m}(2)^{2m}(\sqrt{2\beta}\tau)^{2n}}{(m)!(n-m)!(n-\frac{m}{2})}\Bigg{)} \Bigg{]}
\end{aligned}
\end{equation}
It is useful to note, that one can perform the first part of the sum over $n$ exactly by noting that
\begin{align}
    \sum _{n=1}^{\infty } \frac{2^{\frac{5 n}{2}} \Gamma \left(\frac{n}{2}\right) \left(\sqrt{\beta } \tau \right)^n}{n!}=16 \beta  \tau ^2 \, _2F_2\left(1,1;\frac{3}{2},2;8 \beta  \tau ^2\right)+\pi \Erfi\left(2 \sqrt{2} \sqrt{\beta } \tau \right)
\end{align}
Therefore, the final result is:
\begin{align}
   \chi_{\beta}(\tau) &{}= -e^{S_0}\frac{\tau}{8\pi\beta} \Bigg{[}  \Gamma(0, 2\beta \tau^2)(e^{-8\beta\tau^2} - 1) + e^{-8\beta \tau^2}16 \beta  \tau ^2 \, _2F_2\left(1,1;\frac{3}{2},2;8 \beta  \tau ^2\right) \nonumber \\[5pt] 
   &+e^{-8\beta \tau^2}\pi \Erfi\left( \sqrt{8\beta \tau^2}  \right)- e^{-8\beta \tau^2}\sum_{n=1}^{\infty}\left(\sum_{m=1}^{n}\frac{(-1)^{n+m}(2)^{2m}}{(m)!(n-m)!(n-\frac{m}{2})}\right)\left(2\beta\tau^2\right)^n \Bigg{]} 
   \label{eq:J24}
\end{align}
Using, 
\begin{align}
    \kappa^{\text{GOE}}_{\beta}(\tau) = 2\kappa^{\text{GUE}}_{\beta}(\tau) +\chi_{\beta}(\tau)
\end{align}
with~\eqref{eq:J24} and~\eqref{eq:SFF_GUE} reproduces~\eqref{eq:SFF_GOE} in the main text. This concludes the derivation of the canonical SFF from RMT universality.
\newpage
\bibliography{lib_update}

\end{document}